\pdfoutput=1

\documentclass[iop,revtex4]{emulateapj}
\usepackage{color}
\usepackage{amsmath}
\usepackage{xcolor}
\usepackage{mathrsfs}
\usepackage{natbib}

\shortauthors{Bryan et al.}
\shorttitle{Searching for Scatterers}
\slugcomment{{\sc Accepted to ApJ:} June 16, 2016}

\begin{document}

\title{Searching for Scatterers:  High-Contrast Imaging of Young Stars Hosting Wide-Separation Planetary-Mass Companions}

\author{
Marta L. Bryan\altaffilmark{1},
Brendan P. Bowler\altaffilmark{2,3,4},
Heather A. Knutson\altaffilmark{3},
Adam L. Kraus\altaffilmark{2},
Sasha Hinkley\altaffilmark{5},
Dimitri Mawet\altaffilmark{1},
Eric L. Nielsen\altaffilmark{6,7},
Sarah C. Blunt\altaffilmark{6,8}
}

\altaffiltext{1}{Cahill Center for Astronomy and Astrophysics, California Institute of Technology,
1200 East California Boulevard, MC 249-17, Pasadena, CA 91125, USA}
\altaffiltext{2}{McDonald Observatory and Department of Astronomy, University of Texas at Austin, Austin, TX 78712, USA}
\altaffiltext{3}{Division of Geological and Planetary Sciences, California Institute of Technology, Pasadena, CA 91125 USA}
\altaffiltext{4}{McDonald Prize Fellow}
\altaffiltext{5}{University of Exeter, Physics Department, Stocker Road, Exeter EX4 4QL, UK}
\altaffiltext{6}{SETI Institute, Carl Sagan Center, 189 Bernardo Avenue, Mountain View, CA 94043}
\altaffiltext{7}{Kavli Institute for Particle Astrophysics and Cosmology, Stanford University, Stanford, CA 94305}
\altaffiltext{8}{Department of Physics, Brown University, Providence, RI 02912}

\begin{abstract}

We have conducted an angular differential imaging survey with NIRC2 at Keck in search of close-in substellar companions to a sample of seven systems with confirmed planetary-mass companions (PMCs) on wide orbits ($>$50 AU).  These wide-separation PMCs pose significant challenges to all three possible formation mechanisms:  core accretion plus scattering, disk instability, and turbulent fragmentation.  We explore the possibility that these companions formed closer in and were scattered out to their present-day locations by searching for other massive bodies at smaller separations.  The typical sensitivity for this survey is $\Delta K$ $\sim$ 12.5 at 1$"$.  We identify eight candidate companions, whose masses would reach as low as one Jupiter mass if gravitationally bound.  From our multi-epoch astrometry we determine that seven of these are conclusively background objects, while the eighth near DH Tau is ambiguous and requires additional monitoring.  We rule out the presence of $>$7 M$_{\rm Jup}$ bodies in these systems down to 15 -- 50 AU that could be responsible for scattering.  This result combined with the totality of evidence suggests that dynamical scattering is unlikely to have produced this population of PMCs.   We detect orbital motion from the companions ROXs 42B b and ROXs 12 b, and from this determine 95$\%$ upper limits on the companions' eccentricities of 0.58 and 0.83 respectively.  Finally, we find that the 95$\%$ upper limit on the occurrence rate of additional planets with masses between 5 -- 15 M$_{\rm Jup}$ outside of 40 AU in systems with PMCs is 54$\%$.  

\end{abstract}

\section{Introduction}
Observational studies of exoplanet systems present a unique opportunity to probe the mechanisms behind planet formation. Over the past decade, surveys using a variety of techniques (radial velocity, transit, microlensing, direct imaging) have revealed a multitude of new systems with astoundingly diverse properties. Many of these systems are difficult to explain within the framework of standard planet formation theories \citep[e.g.][]{Pollack1996,Boss2006}, and have forced theorists and observers alike to re-evaluate their narratives for planet formation and migration. Perhaps one of the biggest challenges for planet formation models comes from direct imaging surveys, which have uncovered of a new population of young planetary-mass companions (PMCs) ($<$ 15 M$_{\rm Jup}$) located beyond 50 AU.

In 2004 Chauvin et al. discovered a 5 M$_{\rm Jup}$ companion 2M1207 b orbiting 55 AU away from a 25 M$_{\rm Jup}$ brown dwarf.   Shortly afterwards, additional discoveries of other wide-separation PMCs such as AB Pic b \citep{Chauvin2005}, DH Tau b \citep{Itoh2005}, and CHXR 73 b \citep{Luhman2006} drove observers and theorists to question how this growing population of objects formed \citep{Lodato2005, Boss2006}.   To date, fifteen PMCs at large orbital distances have been confirmed, most of which are extremely young, $<$10 Myr old \citep{Bowler2014}.  Three possible formation routes have been proposed for these wide-separation planets, including direct collapse from molecular cloud fragmentation, disk instability, and core accretion plus gas capture, but all three have significant problems explaining this population of PMCs.
 
In the process of turbulent fragmentation, all stellar and substellar objects begin as opacity-limited fragments with masses of a few Jupiter masses and subsequently begin to accrete gas from the molecular cloud \citep{Low1976}. Hydrodynamical star formation simulations have shown that in order to stop accretion at brown dwarf or planetary masses, PMCs must either form at nearly the same time that the circumstellar envelope is exhausted, or else they must be dynamically ejected from the densest regions of gas before they are able to accrete much additional mass \citep{Bate2002, Bate2009, Bate2012}. This mechanism has a very difficult time producing binaries with the high mass ratios needed to match the observed wide-separation planetary systems. 

In models of disk instability, gas giant planets form rapidly via fragmentation of a gravitationally unstable disk. For this model to work, the disk needs to be massive enough and cold enough to gravitationally collapse. In the majority of scenarios, the disk surface densities beyond 100 AU are too low for gravitational instability to operate. While some models show that disk fragmentation can occur outside 100 AU \citep{Dodson2009, Boss2006, Vorobyov2013}, the fragments rarely survive to become full-fledged planetary embryos.  This low survival probability is due to processes such as inward migration and accretion onto the host star, or ejection from the system due to dynamical interactions.  While it has been suggested that disk instability could be effective for exceptionally massive disks, this is an extremely limited region of disk parameter space \citep{Vorobyov2013}.

Finally, in the core accretion model, cores grow via successions of two-body collisions between solids until they are massive enough to start runaway gas accretion \citep{Pollack1996, Alibert2005}. In situ formation of massive wide-separation planets through core accretion is unlikely since the timescale to grow massive cores at these separations is longer than the observed lifetimes of protoplanetary disks.  However, recent simulations of core formation via pebble accretion have shown that gas giant cores can form at separations out to 50 AU comfortably before the gas in the disk dissipates \citep{Lambrechts2012}.  Furthermore, it might be possible for these giant planets to form closer to the star and be subsequently scattered out beyond 100 AU by another planet in the system. One potential scenario is that if multiple planet-planet scatterings occur, these giant planets could permanently end up in stable, wide-separation orbits \citep{Scharf2009}.  Simulations have shown that in this case, these wide-separation planets have high eccentricities of $>$0.5 \citep{Scharf2009,Nagasawa2011}.  While planet-planet scattering seems to be a potential solution, it requires another body in the system at least as massive as the wide separation planets. 

Thus far, despite hundreds of hours of AO imaging, only one multi-planet system has been confirmed with this technique, HR 8799 \citep{Marois2008, Marois2010}.  Recently, two surveys have found evidence of additional planets in two more systems:  LkCa 15 and HD 100546  \citep{Kraus2012, Quanz2015, Sallum2015, Currie2015}.  Searching for additional planets in these directly imaged systems is critical to understanding the formation and orbital evolution of planets at wide separations, a parameter space currently explored solely by the direct imaging technique.  

In this study, we explore the possibility that the observed wide-separation PMCs formed closer in to their host stars, and were scattered out to their present day locations by another massive companion within the system.  We conducted an angular differential imaging (ADI) survey with NIRC2 at Keck in search of close-in substellar companions to a sample of seven systems with confirmed PMCs on extremely wide orbits.  Our observations are sensitive to companions at significantly lower masses and smaller separations than previous studies of these systems, and allow us to place much stronger constraints on the presence of inner companions.  We also use these same systems to calculate the first estimate of the multiplicity of directly imaged planetary systems.

This paper is structured in the following manner.  In Section 2 we describe the selected sample of systems and the methods for obtaining the ADI imaging data.  In Section 3 we describe the PCA reduction of the images as well as a new method to simultaneously calculate astrometry and relative photometry of candidate companions.  Finally, in Section 4 we discuss our results and their implications for the formation mechanisms of this population of wide-separation PMCs.

\section{Observations}
\subsection{Target Selection}

We selected our targets from the sample of 15 systems with confirmed companions beyond 50 AU with mass ranges that are either below or straddle the deuterium burning limit ($<$ 15 M$_{\rm Jup}$).  These systems are as a whole extremely young, which translates into higher sensitivity to lower mass planets at smaller separations.  From this larger sample, we selected targets that were observable from Keck and that had previously been imaged only with short integrations.  This would allow our deeper follow-up imaging to achieve unprecedented levels of sensitivity in these systems.  Altogether, we targeted seven systems:  ROXs~42B, ROXs~12, HN~Peg, HD~203030, DH~Tau, LP~261--75, and 2MASS~J012250--243950.  Table 1 summarizes the properties of this sample.  In addition, we targeted 2MASS~J162627744--2527247, which does not have a previously confirmed wide-separation PMC.  This star is a wide separation stellar companion to ROXs 12, located $\sim$40" away.\footnote{There was some confusion with regards to follow-up observations of ROXs 12.  The coordinates for ROXs 12 listed in Simbad and in both the discovery and confirmation papers of the PMC ROXs 12b \citep{Ratzka2005, Kraus2014} are for 2M1626-2527, which does not have a confirmed PMC.  The correct coordinates for ROXs 12 are listed in Table 1.}  Not only do the two stars show identical proper motion, but 2M1626--2527 and ROXs 12 also exhibit WISE excesses, indicating that these objects form a wide binary, are disk-bearing, and are young.  

\begin{deluxetable*}{lccccccccc}
\tabletypesize{\scriptsize}
\tablewidth{0pc}
\tablecaption{Target Sample}
\tablehead{
\colhead{System}&
\colhead{RA}   &
\colhead{Dec} &
\colhead{Pri. SpT} &
\colhead{m$_{K}$} & 
\colhead{m$_{R}$} &
\colhead{Dist.} &
\colhead{M$_{\rm comp}$} &
\colhead{Age}  &
\colhead{Ref.}  \\
\colhead{}  &
\colhead{(J2000)}  &
\colhead{(J2000)}  &
\colhead{}  &
\colhead{(mag)} &
\colhead{(mag)}  &
\colhead{(pc)}  &
\colhead{(M$_{\rm Jup}$) } &
\colhead{(Myr)}  &
\colhead{}
}
\startdata
2M0122--2439  & 01 22 50.94  & --24 39 50.6 & M3.5 & $9.20 \pm 0.03$ & $13.6$ & $36 \pm 4$ & 12 -- 25 &  $120 \pm 10$ &  1, 2, 3\\
DH Tau  & 04 29 41.56 & +26 32 58.3 & M1 & $8.18 \pm 0.03$ & $12.1$ & $145 \pm 15$  &   $12^{+10}_{-4}$ &  $1-2$ & 4, 6, 10, 13\\
LP 261--75  & 09 51 04.60 & +35 58 09.8 & M4.5  & $9.69 \pm 0.02$ & $14.4$ & $32.9^{+3}_{-2}$  &   $20^{+10}_{-5}$ & $100-200$   & 1, 6, 7, 10  \\
2M1626--2527  & 16 26 27.75 & --25 27 24.7 & M0 &  $9.21 \pm 0.03$ & $15.8$   & $120 \pm 10$ & $\cdots$ & $8^{+4}_{-3}$   & 2, 4, 6 \\
ROXs 12  & 16 26 28.10 & --25 26 47.1 & M0 & $9.10 \pm 0.03$ & $13.5$ & $120 \pm 10$ & 12 -- 20 & $8^{+4}_{-3}$  &  2, 4, 6, 12\\
ROXs 42B  & 16 31 15.02  & --24 32 43.7 & M1 & $8.67 \pm 0.02$ & $13.4$ & $120 \pm 10$  &6 -- 14 & $7^{+3}_{-2}$   &  2, 4, 5, 6, 8\\
HD 203030  & 21 18 58.22 & +26 13 49.9 & G8 & $6.65 \pm 0.02$ & 7.9 & $40.9 \pm 1.2$ &  $23^{+8}_{-11}$ & $130-400$   &  6, 9, 10\\
HN Peg  & 21 44 31.33 & +14 46 19.0 & G0 & $4.56 \pm 0.04$ & 5.6 & $18.4 \pm 0.3$  &  $21 \pm 9$ & $300-400$   &  6, 10, 11
\enddata
\tablecomments{References: (1) \citet{Bowler2013}, (2) \citet{CMC2011}, (3) \citet{Cutri2013}, (4) \citet{Kraus2014}, (5) \citet{Currie2014}, (6) \citet{Cutri2003}, (7) \citet{Reid2006}, (8) \citet{Zacharias2012}, (9) \citet{Metchev2006}, (10) \citet{Zacharias2005}, (11) \citet{Luhman2007}, (12) \citet{Skiff2013}, (13) \citet{Itoh2005}}
\end{deluxetable*}

\begin{deluxetable*}{lcccccccc}
\tabletypesize{\scriptsize}
\tablewidth{0pc}
\tablecaption{Keck/NIRC2 Observations of PMC Systems}
\tablehead{
\colhead{System}&
\colhead{UT Date}   &
\colhead{Filter} &
\colhead{Coronagraph Diam.} &
\colhead{No. of Exp.}&
\colhead{Tot. Exp. Time}&
\colhead{Rot.} &
\colhead{Airmass \footnotemark[1]} &
\colhead{FWHM \footnotemark[1,2]} \\
\colhead{}  &
\colhead{}  &
\colhead{}  &
\colhead{(mas)}&
\colhead{}  &
\colhead{(min)}  &
\colhead{($\deg$)} &
\colhead{} &
\colhead{(mas)}
}
\startdata

2M0122-2439  &  2014 Nov 9  &  $K_{S}$  & 600 & 30  &  30  &  11.0 & 1.44 & 46.5 $\pm$ 0.6\\
DH Tau &  2014 Dec 7  &  $K_{S}$  & 600& 25 &25 & 57.6  & 1.07& 45.9 $\pm$ 1.9\\  
DH Tau &  2015 Nov 04  &  $K_{S}$  & 600& 25  &  25  &  36.7 &1.01& 46.3 $\pm$ 1.2\\
2M1626--2527 &  2014 May 13  &  $K_{S}$& 600 & 28 & 28 & 11.3 & 1.44& 73.6 $\pm$ 11.6\\
2M1626--2527 & 2015 Jun 23  &  $K_{S}$  & 600& 25 & 25& 9.5 & 1.48&  45.8 $\pm$ 1.5\\
ROXs 12 &  2011 Jun 23  &  $K_{P}$ &300 & 27 & 13.5 & 5.4  & 1.56 & 47.7 $\pm$ 1.6\\
ROXs 12 & 2015 Aug 27  &  $K_{S}$  &600& 20 & 20& 6.4 & 1.54 & 53.0 $\pm$ 9.3\\
ROXs 42B  &  2011 Jun 23  &  $K_{P}$ & 300& 46 & 23 & 13.9  & 1.44& 45.2 $\pm$ 4.9\\
ROXs 42B &  2014 May 13  &  $K_{S}$ &600 &30 &30 & 12.3  & 1.41 & 60.1 $\pm$ 10.1\\
HD 203030 &  2014 Nov 9 &  $K_{S}$ &600 & 60 & 30 & 12.4  &1.06 & 43.0 $\pm$ 0.3\\
HD 203030 &  2015 Jun 3  &  $K_{S}$ &600 & 80 & 40& 80.9  & 1.02&  40.4 $\pm$ 1.6\\
HN Peg &  2014 Aug 4  &  $K_{S}$ & 600& 50 & 25& 102.3  & 1.01& 47.7 $\pm$ 0.7\\
HN Peg &  2015 Jun 2  &  $K_{S}$ & 600& 70 & 35&  32.1  &1.04 &  39.6 $\pm$ 1.2
\enddata
\footnotetext[1]{Values averaged over the total duration of the observations for each target, which typically spanned 20 - 30 minutes.}
\footnotetext[2]{The reported uncertainty on each FWHM value is the standard deviation of the PSF over the duration of the observations.}
\end{deluxetable*}
\subsection{NIRC2 Imaging}
We used the near-infrared imaging camera NIRC2 at the Keck II 10 m telescope for all of our observations.  Adaptive optics imaging was carried out in natural guide star mode using the narrow camera. Due to the realignment of the Keck II AO and NIRC2 system in April 2015, for epochs taken prior to this date we used a plate scale of $0.009952 \pm 0.000002$ arcsec/pixel \citep{Yelda2010}, and for epochs taken after this date we used a plate scale of $0.009971 \pm 0.000004$ arcsec/pixel \citep{Service}.  The field of view of the 1024$\times$1024 array is 10.2"$\times$10.2".  For each system we obtained a total integration time of $\sim 30$ minutes in ADI mode with an average of $\sim 30$ degrees of sky rotation.  For each image, the star was centered behind a coronagraph, which for nearly all images was the 600 mas diameter coronagraph.  This coronagraph is partially transparent with 6.65 +/- 0.10 magnitudes of attenuation in Ks band \citep{Bowler2015}.  Due to high noise levels in the lower left quadrant of the detector, we positioned the occulting spot (already fixed at row 430) at column 616.  When possible, we observed these systems as they were transiting to maximize the rotation achieved during the observation, which makes post-processing PSF subtraction more effective.  The inner working angle achieved for these observations is 300 mas, while the outer working angle for complete FOV coverage is $\sim 4"$.  All observations were taken with the $Ks$ filter, which maximizes the Strehl ratio while avoiding the high sky backgrounds encountered in L band.  We took second epoch data for the systems where we identified a candidate companion.

\section{Analysis}
\subsection{PCA Image Reduction}

After removing bad pixels and cosmic rays and flat-fielding, we applied the \citet{Yelda2010} distortion correction to raw images taken before April 2015, and applied the updated \citet{Service} distortion solution to later epochs.  We then used principle component analysis (PCA) to further reduce these images.  PCA is an algorithm that has recently been applied to high contrast imaging for increasing the contrast achievable next to a bright star.  In short, PCA is a mathematical technique that relies on the assumption that every image in a stack can be represented as a linear combination of its principle orthogonal components, selecting structures that are present in most of the images.  The stellar PSF, composed of a sum or orthogonal components, is subtracted from each image, thereby providing access to faint companions at contrasts below the speckle noise.  We used a PCA routine following the method presented in \citet{Soummer2012} which uses the KLIP algorithm.  

The optimal number of principle components to use in a reduction is set by the trade-off between speckle noise and self-attenuation of the signal of interest.  Too few components might not subtract enough speckle noise near the star, and too many may lead to self-subtraction of the planetary signal, reducing the achievable contrasts.  In our analysis, we optimized the number of components used for each individual system empirically.  We reduced the data for each system with different numbers of principle components, then compared resulting contrast curves that were calibrated for self-attenuation by injecting fake companions.  For each system we adopted the number of principle components that corresponded to the most favorable contrast as the optimal number.  These ranged from 5 -- 20 principle components for systems in our sample.  

We found a total of nine candidate companions at a wide variety of separations in the eight systems that we observed.  These candidate companions are shown in Figure 1, and the contrast curves for all systems observed are shown in Figure 2.  We determine our contrast curves by calculating the noise level in our images as a function of radial distance using the standard deviation on concentric annuli of width FWHM of the stellar PSF.  The noise level at each radius is corrected for self-subtraction by dividing by the self-attenuation at that radius.  This self-attenuation factor is calculated by injection and recovery of sources with known magnitudes at different radii.  We present the 5$\sigma$ contrast curves in Figure 2, which are simply our noise levels divided by the self-attenuation factor multiplied by a factor of 5.  We list 5$\sigma$ contrast values for a range of angular separations for each target in Table 3.  We note that these contrasts are often limited by small PA rotation and subsequent ADI self-subtraction.

\begin{figure*}
\includegraphics[width=1\textwidth]{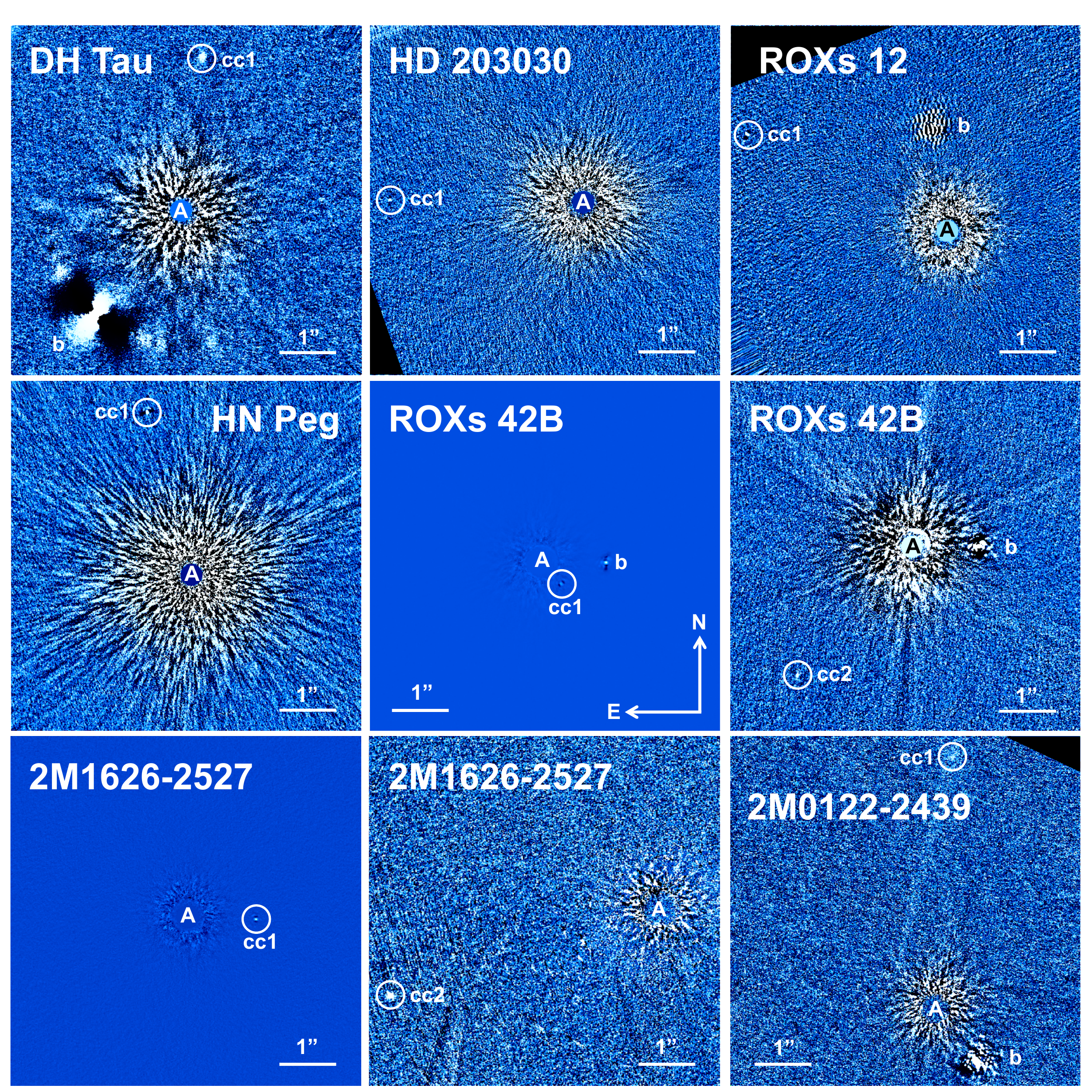}
\caption{Candidate companions in our sample.  All images are north-aligned.  ROXs 42B and 2M1626-2527 are shown twice with different stretches to accommodate candidate companions with significantly different flux ratios.  Some of the known companions exhibit speckle-like features in these images.  This is due to the fact that these bright companions were not masked during the PCA reduction, so some of the PCA components were structured to subtract away the signal of the confirmed companions.  The PCA algorithm was able to more successfully subtract away companions with small amounts of rotation (i.e. ROXs 12 b) in comparison to companions with large amounts of rotation (i.e. DH Tau b).}
\end{figure*}

\begin{deluxetable*}{lccccccc}
\tabletypesize{\scriptsize}
\tablewidth{0pc}
\tablecaption{Contrast Curves}
\tablehead{
\colhead{System}&
\colhead{0.5"} &
\colhead{1"}  &
\colhead{1.5"}&
\colhead{2"}&
\colhead{2.5"}&
\colhead{3"}&
\colhead{3.5"}
}
\startdata
2M0122-2439  &  $1.4\times10^{-3}$ &  2.5$\times10^{-5}$  &  1.0$\times10^{-5}$  &  $5.9\times10^{-6}$  &  $5.6\times10^{-6}$  &  $5.1\times10^{-6}$  &  $4.8\times10^{-6} $\\
DH Tau &  $2.7\times10^{-4}$  &  $3.2\times10^{-5}$  &  $9.1\times10^{-6}$  & $7.4\times10^{-6}$  &  $6.9\times10^{-6}$  &  $5.2\times10^{-6}$  &  $5.1\times10^{-6}$ \\   
LP261-75  &  $1.6\times10^-{4}$  &  $2.2\times10^{-5}$  &  $1.5\times10^{-5}$  & $1.6\times10^{-5}$  &  $1.5\times10^{-5}$ &  $1.4\times10^{-5}$  &$1.5\times10^{-5}$ \\
2M1626--2527 &  $1.6\times10^{-3}$  &  $8.4\times10^{-5}$  &  $2.7\times10^{-5}$  &  $1.6\times10^{-5}$ & $1.7\times10^{-5}$  &  $1.5\times10^{-5}$  &  $1.5\times10^{-5}$\\
ROXs 12 &  $2.7\times10^{-2}$  &  $3.0\times10^{-4}$  &  $2.9\times10^{-5}$  &  $2.0\times10^{-5}$  &  $1.5\times10^{-5}$  &  $1.2\times10^{-5}$  &  $1.3\times10^{-5}$ \\
ROXs 42B  &  $5.1\times10^{-4}$  &  $3.6\times10^{-5}$  &  $1.1\times10^{-5}$  &  $5.8\times10^{-6}$  &  $5.1\times10^{-6}$  &  $4.7\times10^{-6}$  &  $4.8\times10^{-6}$ \\
HD 203030 &  $1.2\times10^{-4}$  &  $1.0\times10^{-5}$  &  $2.4\times10^{-6}$  &  $9.7\times10^{-7}$  &  $6.8\times10^{-7}$  &  $6.2\times10^{-7}$  &  $4.9\times10^{-7}$  \\
HN Peg &  $3.4\times10^{-4}$  &  $1.4\times10^{-5}$  &  $2.4\times10^{-6}$  &  $8.2\times10^{-7}$  &  $4.2\times10^{-7}$  &  $3.5\times10^{-7}$  &  $2.9\times10^{-7}$

\enddata
\tablecomments{All of these contrasts correspond to images taken using the $K_S$ filter.}

\end{deluxetable*}

\begin{figure*}
\includegraphics[width=1\textwidth]{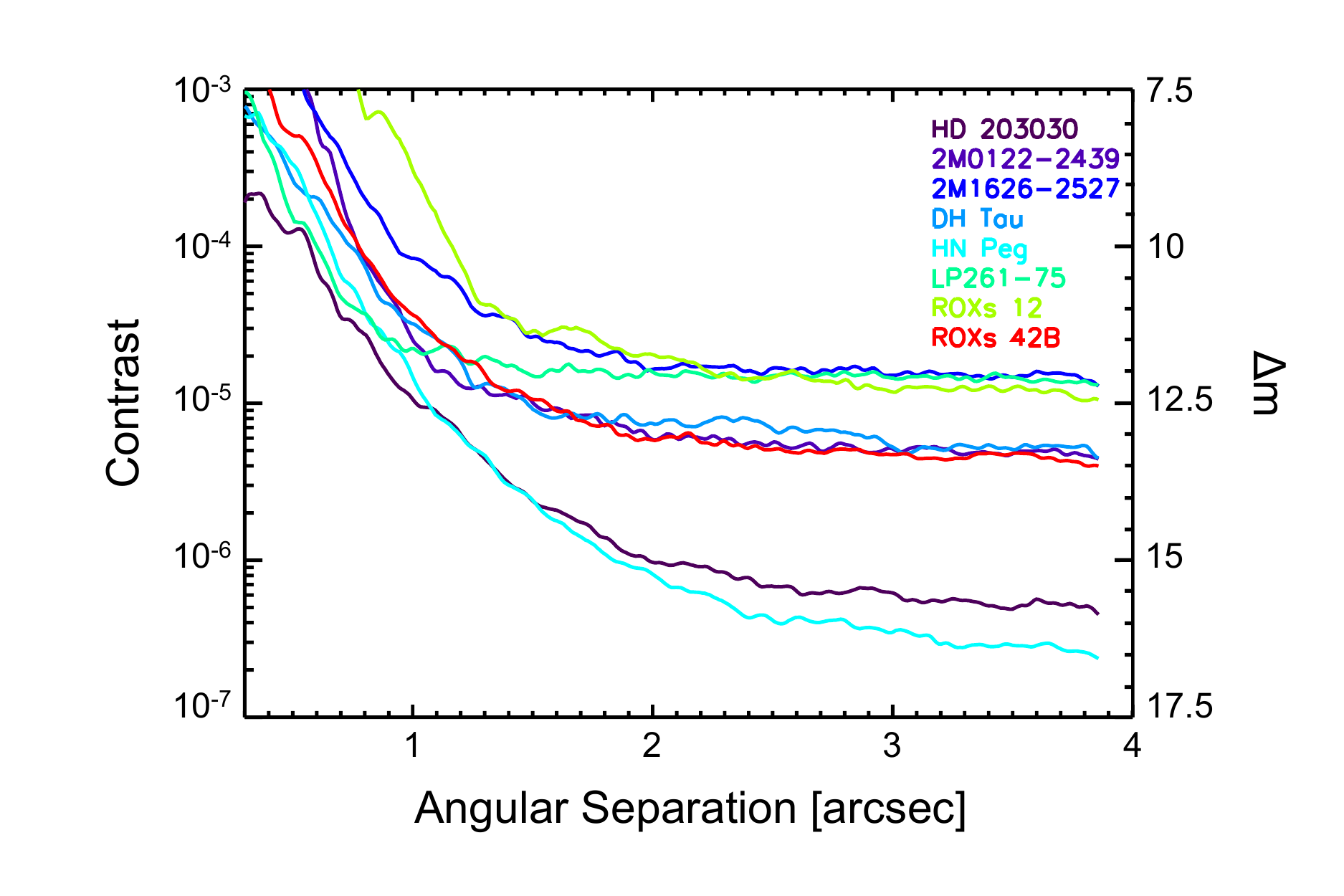}
\caption{$5\sigma$ contrast curves for systems in our sample.  The stellar flux was modified by a throughput correction \citep{Bowler2013} due to attenuation from the coronagraph spot.  The curves have been corrected for self-attenuation of the target of interest using a robust injection and recovery technique.  Note that the flattening of the contrast curves indicates a background-limited regime, as opposed to speckle limitations.}
\end{figure*}

\subsection{Simultaneous Astrometry and Relative Photometry}

Using second epoch data, we can determine whether or not these candidate companions are co-moving.  Typical methods for determining the astrometry and photometry of candidate companions use the final post-processed images for these calculations.  However, we note that because of self-subtraction, using the final LOCI or PCA images to calculate separations, position angles, and their uncertainties can lead to significant biases in the corresponding photometry and astrometry \citep[e.g.][]{Marois2010}.  To avoid this, we developed an MCMC algorithm that simultaneously calculates the astrometry and relative photometry of these candidate companions.  For each iteration in the MCMC process, we injected a negative PSF into each individual science image in the vicinity of the companion of interest prior to de-rotation, where we modeled these negative PSFs as Moffat distributions.  There were three parameters that we varied with each step during the MCMC routine, namely the negative PSF amplitude, separation, and position angle.  We fixed all other free parameters to the values determined from fitting the Moffat distribution to the stellar PSF.  The science images with injected negative PSFs were then run through the PCA reduction routine.  The smaller the RMS noise at the location of the candidate companion, the better the fit of the negative PSF.  The result of this MCMC analysis is a posterior distribution for the amplitude, separation, and position angle of the candidate companions, an example of which is shown in Figure 3.  Figure 4 compares a reduced image with and without the best-fit negative PSF injected at the best fit separation and position angle of the candidate companion.  

\begin{figure*}
\includegraphics[width=1\textwidth]{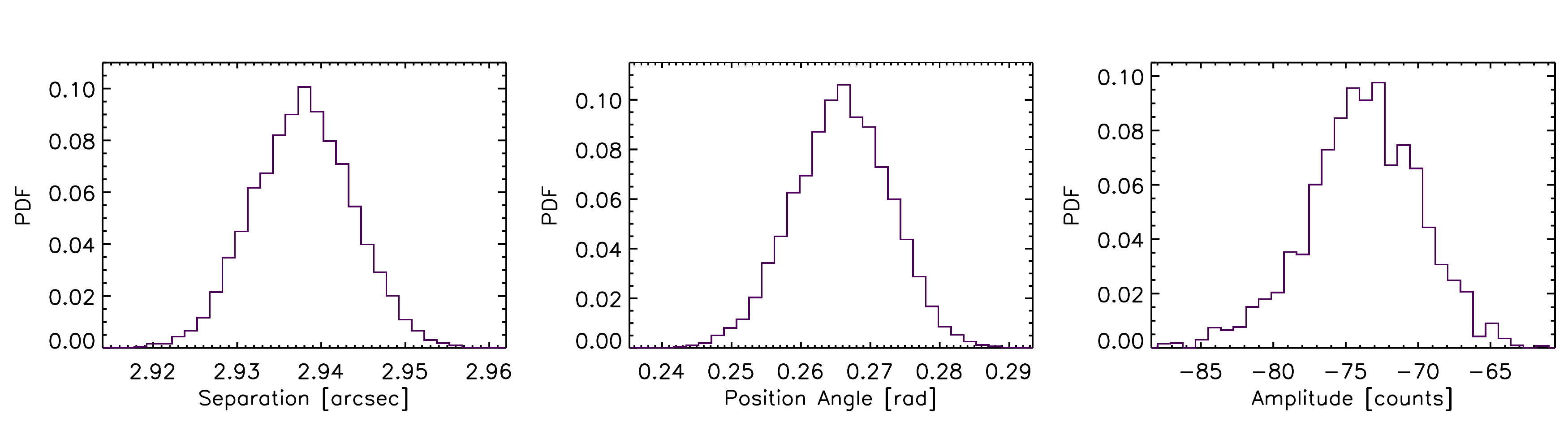}
\caption{Example posterior PDFs from the MCMC astrometry calculation for the candidate companion in our observations of HN Peg.  Note that the amplitude PDF shows the amplitude of the negative PSF injected into the science images.}
\end{figure*}

\begin{figure}
\includegraphics[width=0.5\textwidth]{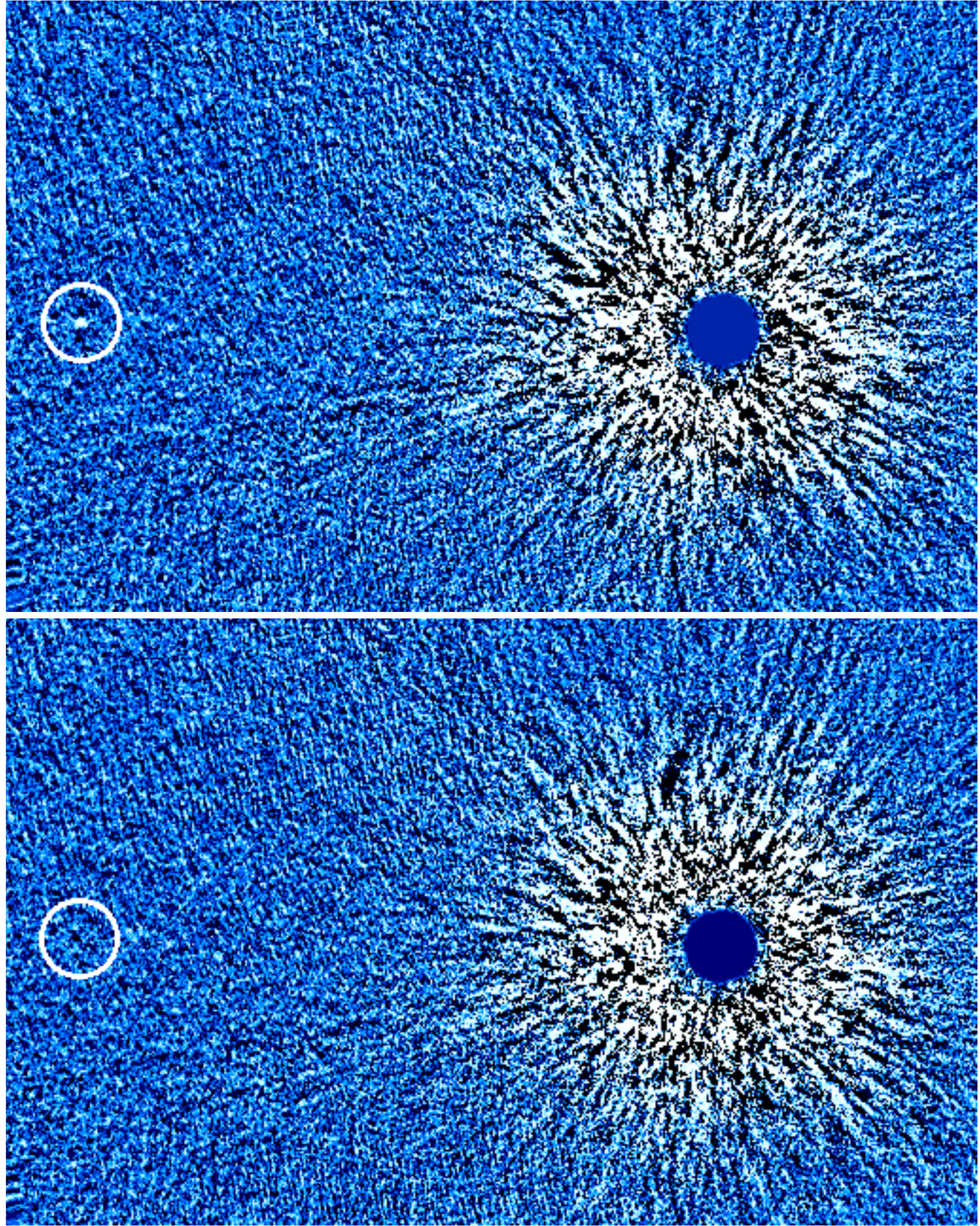}
\caption{Comparison of PCA images of the HD 203030 candidate companion without (top) and with (bottom) the best fit negative PSF injected.}
\end{figure}

In addition to the errors from the candidate companion PDFs for the separation and position angle, after the MCMC program is finished we also account for uncertainties from the registration of the stellar position in each science image, uncertainties from the distortion correction, and uncertainties associated with the plate scale.  We also include the $+0.252$ degree correction for north alignment to the NIRC2 header position angles \citep{Yelda2010} for epochs taken before April 2015, and for subsequent epochs include the $+0.262$ degree correction for north alignment \citep{Service}.  This MCMC technique calculates robust uncertainties from the posterior distributions, without the systematic uncertainties that occur when the reduced LOCI or PCA images are used.  The best-fit separations and position angles for each candidate companion, as well as confirmed companions in these systems, are presented in Table 4.  

While the separation, position angle, and amplitude are the three parameters that were actively varied with each step in the MCMC program, we also track the total flux of the candidate companion.  For each link in the chain, we place an aperture at the separation and position angle of that step and sum the number of counts in the aperture.  While the size of the aperture for a given companion remains fixed, the aperture size ranged from 4 -- 6 pixels depending on the candidate companion FWHM.  The aperture moves with the changes in position angle and separation as the MCMC program progresses, producing a posterior distribution of counts for the candidate companion.  We note that since we calculate the astrometry and relative photometry simultaneously using this MCMC program, we use the same aperture sizes for both of these steps for a given candidate companion.

We next calculate the contrast relative to the host star, $\Delta m$, for each candidate companion.  To determine the flux from the star, the throughput of the occulting spot ($0.0022 \pm 0.0002$, Bowler et al 2015), and the sky noise need to be accounted for.  The measured counts are a combination of the flux from the star plus the flux from the sky, both attenuated by the throughput of the mask.  We denote this combined and attenuated star plus sky flux as $F_{b,\star}$.  The corrected flux for the star is shown in Equation 1. $T$ is the throughput (0.0022) of the mask in $K_{S}$, $F_{sky}$ is the sky flux, and $F_{\star}$ is the flux from the star corrected for both throughput losses and sky noise.  

\begin{equation}
F_{\star} = \frac{F_{\star,b}}{T} - F_{sky}
\end{equation}

\noindent This calculation is performed for each image in the stack; for a stack of $N$ images, there are $N$ values of $F_{\star}$.  The error on this flux value is:

\begin{equation}
\sigma_{F, \star} = \sqrt{\bigg(\frac{F_{b,\star}}{T}\bigg)^2 \times \bigg[\bigg(\frac{\sigma_{\star,b}}{F_{\star,b}}\bigg)^2 + \bigg(\frac{\sigma_{T}}{T}\bigg)^2\bigg] + \sigma_{sky}^2}  .
\end{equation}

\noindent Here, $\sigma_{\star,b}$ is the standard deviation of the $F_{\star,b}$ values for each image in the stack, $\sigma_T$ is the measured error on the throughput, and $\sigma_{sky}$ is the standard deviation of the sky values calculated for each image.

We obtain the companion flux and its uncertainty from the posterior distribution generated from the MCMC analysis, and subtract off sky noise.  The flux ratio between the star and the companion is simply $F_{ratio} = \frac{F_{\star}}{F_{\rm comp}}$, and the error on this flux ratio can be propagated analytically:
\begin{equation}
\sigma_{fr} = F_{ratio} \times \sqrt{\bigg(\frac{\sigma_{F,\star}}{F_{\star}}\bigg)^2 + \bigg(\frac{\sigma_{comp}}{F_{comp}}\bigg)^2}  .
\end{equation}

\noindent The contrast ratio $\Delta m$ in magnitudes is:

\begin{equation}
\Delta m = -2.5\log_{10}{F_{ratio}} .
\end{equation}

Finally, the error on $\Delta m$ is given by:

\begin{equation}
\sigma_{\Delta m} = \frac{2.5}{\ln 10} \times \frac{\sigma_{fr}}{F_{ratio}}  .
\end{equation}

\noindent We present the $\Delta m$ values for each candidate and confirmed companion in Table 4.  We note that because 2M0122-2439 cc1 is so faint we were unable to use our MCMC analysis to calculate the astrometry of that companion (the MCMC chains failed to converge).  Instead, we used centroiding on the final image to obtain the separation and position angle of the candidate companion, and adopted the robust errors calculated for the faint HD 203030 cc1.  Furthermore, since the candidate companion near 2M0122-2439 appears to be extended (its FWHM is about twice that of the stellar PSF), we conclude that is likely a background galaxy, not a bound planet, and exclude it from the rest of the analysis.

\begin{figure*}
\includegraphics[width=1\textwidth]{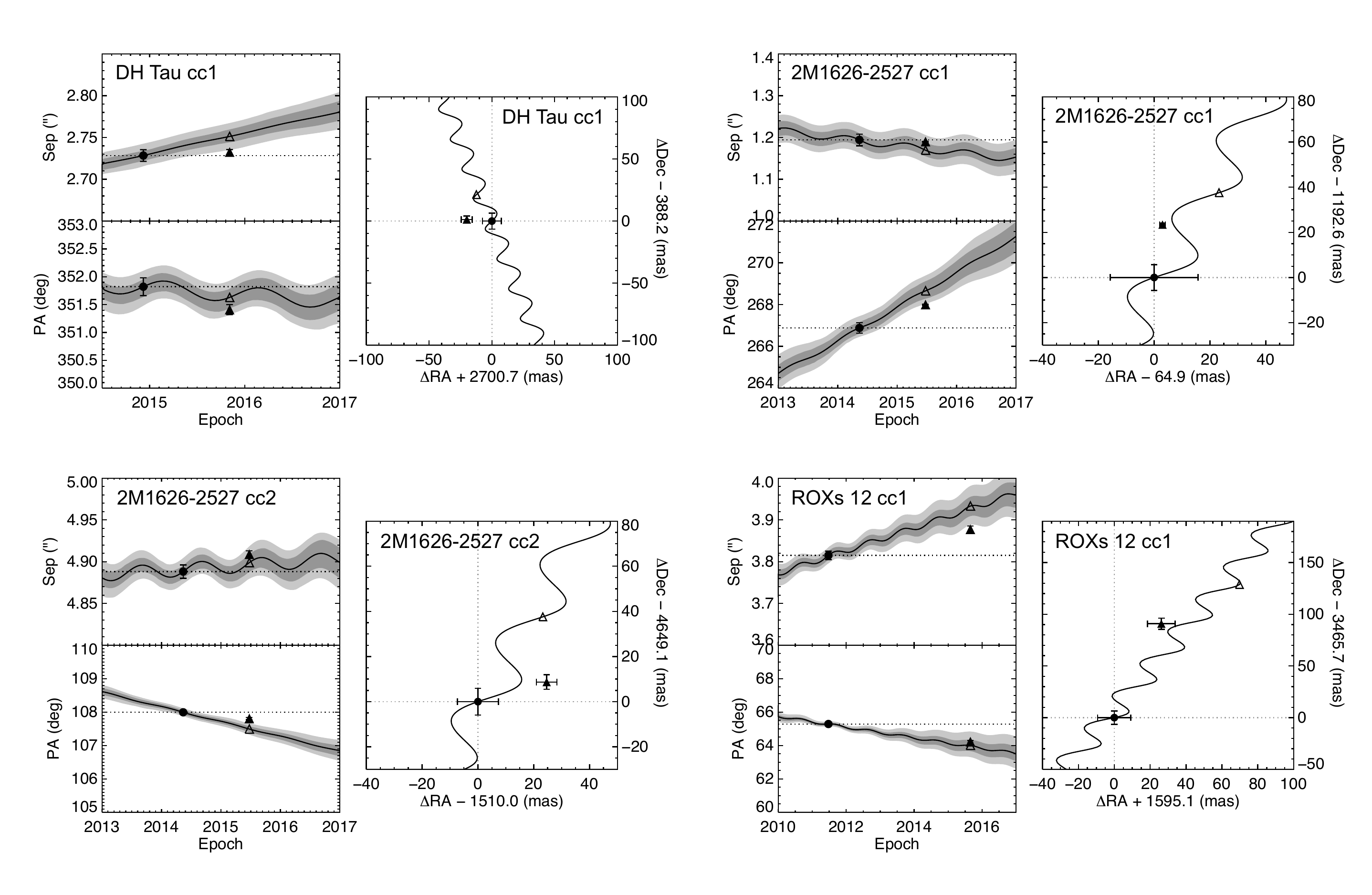}
\caption{These plots show how the candidate companion's astrometry compares to expected trajectories of a co-moving object and a stationary background object.  The first epoch of astrometry is denoted by a filled circle, and the second epoch is denoted by a filled triangle.  The open triangles denote the expected astrometry of a stationary background object at the second epoch.  The dark and light grey regions represent the one and two sigma errors on the predicted background tracks, respectively.  These errors include uncertainties in the distance to the system, proper motion, and astrometry from the reference epoch.  If the candidate companion was bound to the star, the second epoch triangles would fall on the horizontal dotted line (separation and position angle would not change as a function of time, except due to orbital motion).  Top left:  DH Tau candidate companion (cc) 1.  Top right:  2M1626--2527 cc1.  Bottom left:  2M1626--2527 cc2.  Bottom right:  ROXs 12 cc1.}
\end{figure*}

\begin{figure*}
\includegraphics[width=1\textwidth]{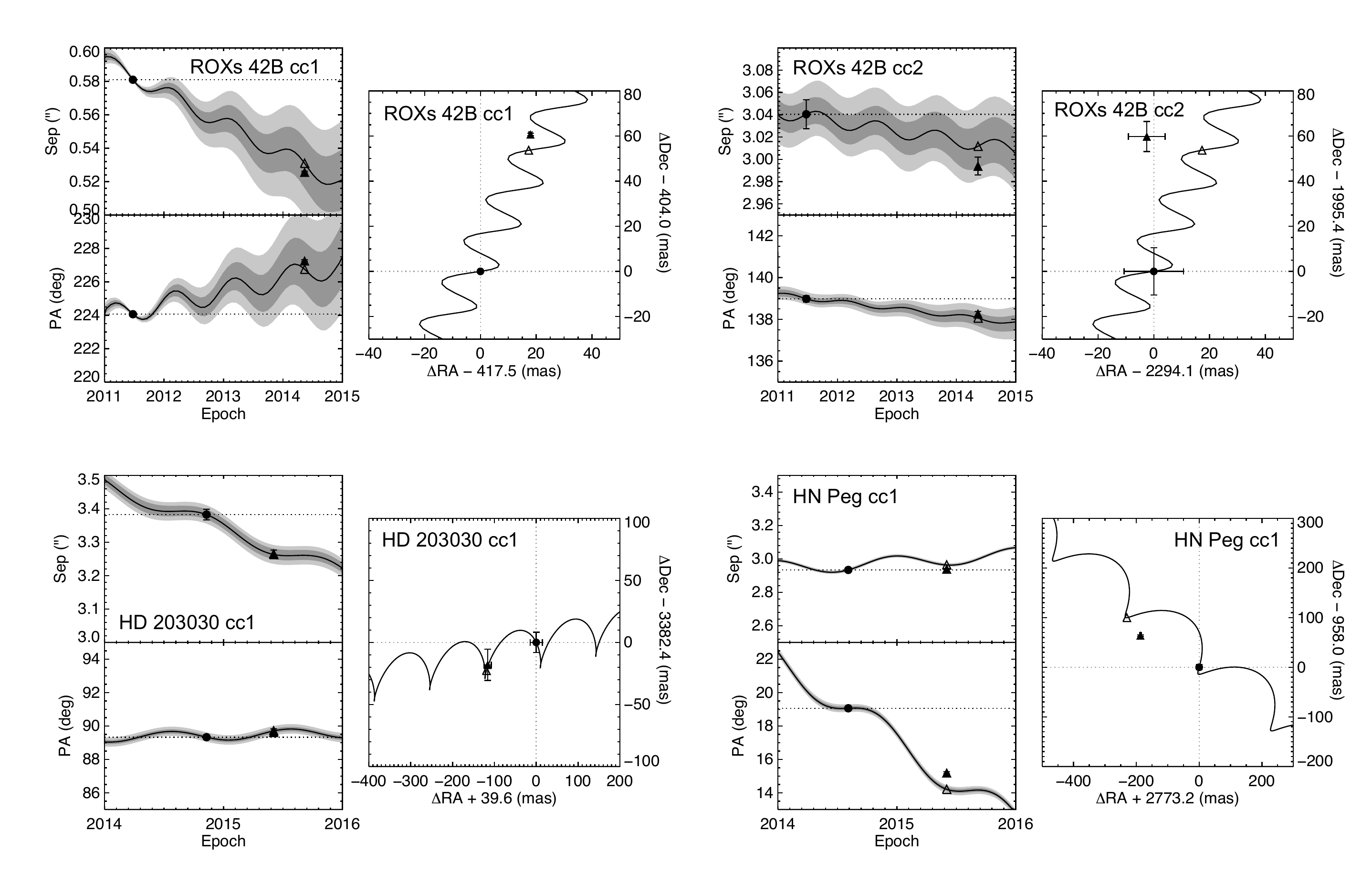}
\caption{Background track plots for four of our candidate companions.  See Figure 5 for more details.  Top left:  ROXs 42B cc1.  Top right:  ROXs 42B cc2.  Bottom left:  HD 203030 cc1.  Bottom right:  HN Peg cc1.}
\end{figure*}

\begin{figure*}
\includegraphics[width=1\textwidth]{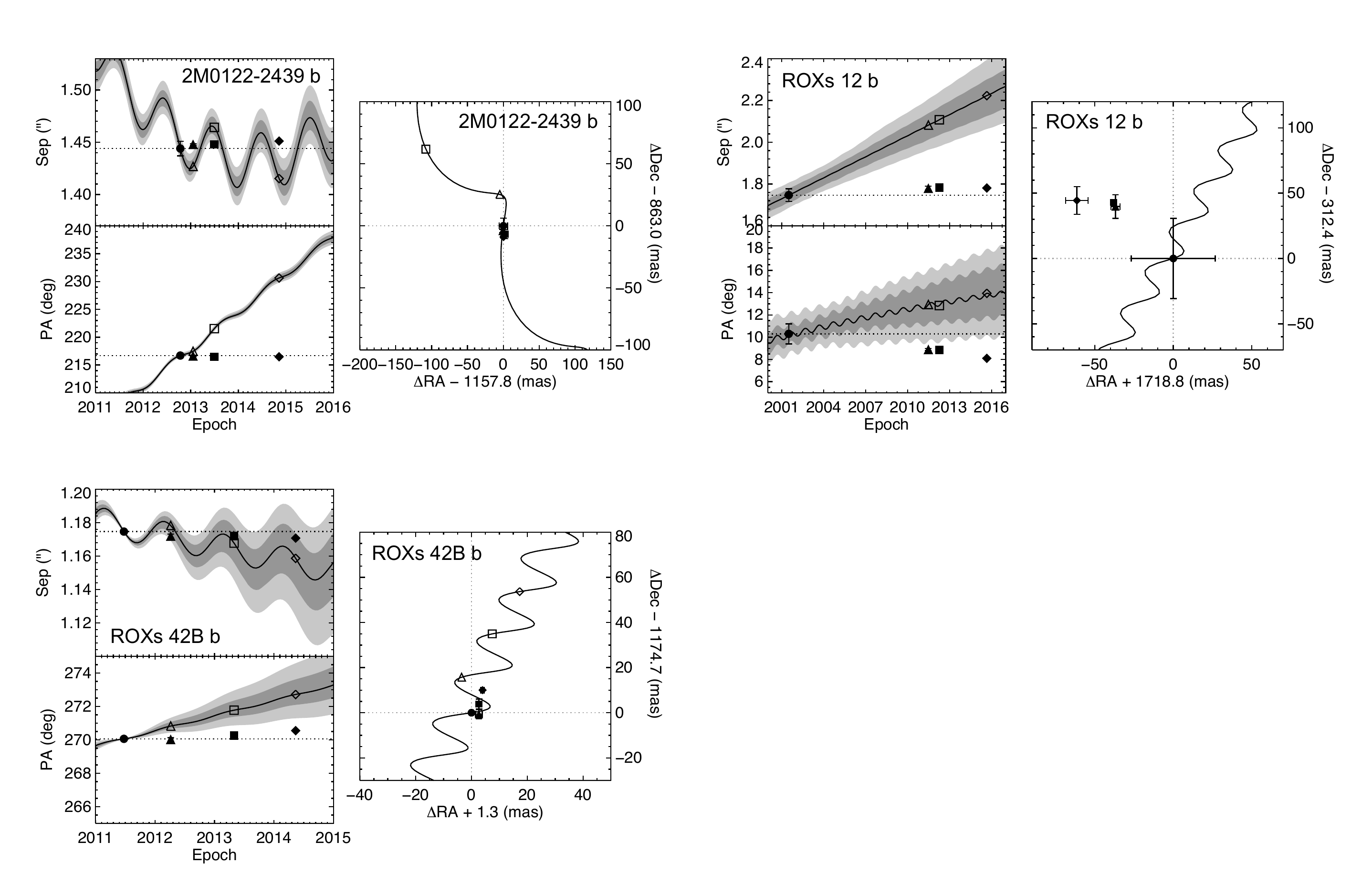}
\caption{These background track plots show the astrometry of three previously confirmed companions in our sample.  Top left:  2M0122--2439 b.  Top right:  ROXs 12 b.  Bottom left:  ROXs 42B b.  In the 2M0122--2439 plot, we include two additional epochs of data in 2012 and 2013 from Bowler et al 2013.  In the ROXs 12 plot, we include two additional epochs, in 2001 from Ratzka et al 2005, and in 2012 from Kraus et al 2014.  In the ROXs 42B plot, we include two additional epochs, in 2001 from Ratzka et al 2005, and in 2012 from Kraus et al 2014. The plots for ROXs 12 b and ROXs 42B b show evidence of orbital motion.  See section 3.3 for details.}
\end{figure*}

\begin{deluxetable*}{lcccccc}
\tabletypesize{\scriptsize}
\tablewidth{0pc}
\tablecaption{Candidate PMC Astrometry and Photometry}
\tablehead{
\colhead{System}&
\colhead{Epoch}   &
\colhead{Filter} &
\colhead{$\rho$ (mas)}&
\colhead{P.A. (deg)} & 
\colhead{$\Delta$m (mag)}  &
\colhead{Sep. (AU)} 
}
\startdata
2M0122--2439 b &  2014.8575  &  $K_{S}$  & $1450^{+1}_{-1}$ & $216.48^{+0.02}_{-0.02}$ &  $4.79 \pm 0.11$  &   52 \\
2M0122--2439 cc1  &  2014.8575  &  $K_{S}$  &  $5238^{+12}_{-16}$  &  $355.75^{+0.14}_{-0.14}$  &  $9.60 \pm 0.20$  &  188\\
DH Tau cc1 &  2014.9342  &  $K_{S}$  &  $2726^{+7}_{-7}$  &  $351.82^{+0.16}_{-0.16}$  &  $8.66 \pm 0.25$  & 395 \\   
DH Tau cc1  &  2015.8438  &  $K_{S}$   &  $2734^{+4}_{-4}$  & $351.35^{+0.10}_{-0.10}$  & $8.45 \pm 0.23$   & 396  \\
DH Tau b  &  2014.9342  &  $K_{S}$   &  $2343^{+1}_{-1}$ &$140.25^{+0.02}_{-0.02}$  &  $5.91 \pm 0.20$    &  340 \\
DH Tau b &  2015.8438  &  $K_{S}$   &  $2339^{+1}_{-1}$  &  $139.94^{+0.02}_{-0.02}$  &  $5.72 \pm 0.28$   &  340\\
2M1626--2527 cc1 &  2014.3644  &  $K_{S}$   &  $1193^{+11}_{-14}$  &  $266.88^{+0.26}_{-0.21}$  &  $7.36 \pm 0.23$   & 143 \\
2M1626--2527 cc1 & 2015.4767  &  $K_{S}$   &  $1189^{+2}_{-1}$  &  $267.93^{+0.03}_{-0.02}$  &  $6.93 \pm 0.13$   &  143  \\
2M1626--2527 cc2 &  2014.3644  &  $K_{S}$  &  $4883^{+8}_{-7}$  &  $107.99^{+0.06}_{-0.06}$  & $7.62 \pm 0.25$   & 586 \\
2M1626--2527 cc2 & 2015.4767  &  $K_{S}$  &  $4901^{+4}_{-3}$  &  $107.82^{+0.03}_{-0.03}$  &  $7.11 \pm 0.13$   &  589 \\
ROXs 12 cc1 &  2011.4767  &  $K_{P}$    &  $3811^{+10}_{-10}$  &  $65.29^{+0.08}_{-0.07}$  &  $\dots$  &  457 \\
ROXs 12 cc1 & 2015.6548  &  $K_{S}$  &  $3877^{+15}_{-14}$  &  $64.12^{+0.09}_{-0.09}$  &  $8.51 \pm 0.13$   & 465  \\
ROXs 12 b  &  2011.4767  &  $K_{P}$  &  $1778^{+1}_{-1}$ & $8.90^{+0.08}_{-0.08}$ &  $\dots$  &  213 \\
ROXs 12 b  &  2015.6548  &  $K_{S}$   &  $1786^{+1}_{-1}$ & $8.18^{+0.29}_{-0.30}$ & $4.30 \pm 0.13$    & 214  \\
ROXs 42B cc1  &  2011.4767  &  $K_{P}$  &  $ 580^{+1}_{-1} $&  $224.06^{+0.06}_{-0.06}$  & $\dots$   &  70  \\
ROXs 42B cc1 &  2014.3644  &  $K_{S}$  &  $525^{+1}_{-1}$  &  $227.27^{+0.06}_{-0.06}$  &  $6.23 \pm 0.27$   & 63 \\ 
ROXs 42B cc2  &  2011.4767  &  $K_{P}$  &  $3037^{+14}_{-11}$  &  $138.99^{+0.14}_{-0.15}$  &  $\dots$  &  365  \\ 
ROXs 42B cc2 &  2014.3644  &  $K_{S}$   &  $2991^{+8}_{-7}$  &  $138.27^{+0.10}_{-0.12}$  &  $8.37 \pm 0.25$   &  359  \\
ROXs 42B b  &  2011.4767  &  $K_{P}$  &  $1173^{+1}_{-1}$ & $270.06^{+0.01}_{-0.01}$ &  $\dots$  &  141 \\
ROXs 42B b &  2014.3644  &  $K_{S}$  &  $1170^{+1}_{-1}$  & $270.55^{+0.01}_{-0.01}$ &  $6.16 \pm 0.35$    &  140 \\
HD 203030 cc1 &  2014.8575 &  $K_{S}$  &  $3379^{+12}_{-16}$  &  $89.33^{+0.14}_{-0.14}$  &  $11.17 \pm 0.15$   &  139  \\
HD 203030 cc1 &  2015.4219  &  $K_{S}$  &  $3263^{+10}_{-7}$  &  $89.76^{+0.16}_{-0.22}$  &  $11.46 \pm 0.20$    & 134 \\
HN Peg cc1 &  2014.5918  &  $K_{S}$   &   $2931^{+4}_{-5}$ &   $19.06^{+0.13}_{-0.11}$  &  $12.63 \pm 0.6$    &  54  \\
HN Peg cc1 &  2015.4192  &  $K_{S}$  &  $2933^{+2}_{-2}$ &  $15.22^{+0.05}_{-0.05}$  &  $12.13 \pm 0.48$   & 54 
\enddata
\tablecomments{We do not list $\Delta m$ for the 2011 epochs because we did not have throughput measurements for the 300 mas coronagraph is $K_P$.  Uncertainties in the astrometry and distance estimates to these systems typically lead to errors in separation in AU of 5 - 40 AU.}
\end{deluxetable*}

Our astrometry conclusively shows that seven of the remaining eight candidate companions are background objects, while the nature of the candidate companion near DH Tau is ambiguous.  Figures 5 and 6 show the relative astrometry of each candidate companion compared to the expected background track of a stationary object.   The candidate companion background track plots clearly show that the second epoch astrometry falls on or near the predicted track for a stationary background object.  While some of the second epoch astrometry measurements don't fall precisely on the expected track of a stationary background object, we note that this could be due to small errors in proper motion or distance, which would affect the predicted trajectory of a distant stationary object.  We note that ROXs 42B cc1 was previously identified as likely a background object in the literature \citep{Kraus2014, Currie2014} but our astrometry conclusively shows that it is a background object.  For DH Tau cc1, while the second epoch astrometry falls close to the co-moving line, uncertainties on the expected trajectory of a background object make comovement ambiguous.  The separation of a stationary object at the second epoch differs by $\sim 2.9\sigma$ from the separation we find for DH Tau cc1.  \citet{Zhou2014} published $HST$ UVIS optical photometry for the DH Tau system but they did not report a detection of our candidate companion.  They presented detection limits in both $i$ and $z$ filters, which we can use to place limits on the colors of DH Tau cc1.  We find that the bluest DH Tau cc1 could be is $i$ -- $K$ = 7.8 mag.  Furthermore, line of sight visual extinction is low, 0.0 - 1.5 mag \citep{Strom1989, White2001}.  This apparently red color further motivates additional follow-up for the potentially bound DH Tau cc1.  A third epoch taken when DH Tau is next observable end of 2016 would conclusively determine whether or not DH Tau cc1 is a bound object.  

We also plot the relative astrometry for the previously confirmed companions to ROXs 12, ROXs 42B, and 2M0122-2439 in Figure 7.  We have included astrometry from the literature in addition to the data presented in this paper.  These plots show that follow-up astrometry generally fall near the dotted line denoting co-moving objects.  We do not plot the relative astrometry for DH Tau b, since there are significant systematic offsets for measurements of the companion position angle and separation amongst previous epochs spanning 1999 through 2013.  Table 5 lists literature astrometry measurements that we used for the confirmed companions in each of these three systems.

\begin{deluxetable}{lcccc}
\tabletypesize{\scriptsize}
\tablewidth{0pc}
\tablecaption{Literature Measurements of Confirmed PMC Astrometry}
\tablehead{
\colhead{Companion}&
\colhead{Epoch}   &
\colhead{$\rho$ (mas)}&
\colhead{P.A. (deg)} &
\colhead{ref.}
}
\startdata
2M0122-2439 b  &  2012.7808  &  $1444 \pm 7$  &  $216.7 \pm 0.2$  &  3\\
2M0122-2439 b &  2013.0493  &  $1448.6 \pm 0.6 $  &  $216.4 \pm 0.08$  &  3\\
2M0122-2439 b &  2013.4959  &  $1448 \pm 4$  &  $216.47 \pm 0.07$  &  3 \\
2M0122-2439 b  &  2013.6258  &  $1488 \pm 3$  &  $216.52 \pm 0.09$  &  5\\
ROXs 12 b  &  2001.5014  &  $1747 \pm 30$  &  $10.3 \pm 0.9$  &  2\\
ROXs 12 b  &  2012.2575  &  $1783.0 \pm 1.8$  &  $8.85 \pm 0.06$  &  1\\
ROXs 42B b & 2001.5014 &  $1137 \pm 30$  &  $268.0 \pm 1.5$  &  2 \\
ROXs 42B b & 2005.2904  &  $1157 \pm 10$  &  $268.8 \pm 0.6$  &  4 \\
ROXs 42B b  & 2008.5479  &  $1160 \pm 10$  &  $269.7 \pm 1.0$  &  4 \\
ROXs 42B b  & 2012.2575 &  $1172.0 \pm 1.2$  &  $270.03 \pm 0.10$  &  1  \\
ROXs 42B b  &  2013.3233  &  $1172.5 \pm 1.2$  &  $270.25 \pm 0.10$  &  1
\enddata
\tablecomments{References:  (1) Kraus et al 2014, (2) Ratzka et al 2005, (3) Bowler et al 2013, (4) Currie et al 2014, (5) Bowler et al 2015}
\end{deluxetable}

\subsection{Orbital Motion}

We tested ROXs 42B b, 2M0122-2439 b, and ROXs 12b for evidence of orbital motion.  Assuming a face on, circular orbit, we find that between the first and last epoch of ROXs~42B~b, the maximum amount of change we would expect to see in position angle is 0.6 degrees.  The actual change in PA between the first and last epochs is $0.49 \pm 0.02$ degrees.  We performed a linear fit to all epochs with uncertainties in PA and in separation and compared these to the best-fit constants using evidence ratios.  Evidence ratios use Akaike's information criterion (AIC) to quantitatively compare models.  They are equal to the ratio of each model's Akaike weights, which are a measure of the strength of evidence for a model.  An evidence ratio of 9 comparing model 1 to model 2 would mean that model 1 is 9 times more likely than model 2 given the data.  We label linear fits as preferred if the slope of the line differs from zero by 2 - 4$\sigma$, and highly preferred if this slope is $>4\sigma$ away.

In PA we find the evidence ratio comparing a linear to constant fit for ROXs~42B~b to be $>10^4$, and in separation the evidence ratio is 62.  The best fit slope of the linear fit in PA is $0.1703 \pm 0.0049$ deg/yr, and in separation is $-0.00132 \pm 0.00029$ arcsec/yr.  We therefore conclude that the linear fits are highly preferred, suggesting that the displacements that we see in PA and separation over time are due to the orbital motion of ROXS~42B~b.  

For the confirmed companion orbiting 2M0122-2439, at a separation of only 52 AU we would expect this companion to have moved by 1.3 degrees in PA between the first and last epochs assuming a circular, face-on orbit.  However, we only find a change in PA of $0.2 \pm 0.2$ degrees between the first and last epochs.  Given that the change in PA is consistent with zero, and the evidence ratio for the separation of the companion favors a constant over a linear fit, we conclude that we do not find evidence of orbital motion for 2M0122-2439 b.

Finally, we assess whether orbital motion is evident for the confirmed companion ROXs 12 b.  Between the first and last epochs, assuming a face-on circular orbit we would expect to see a change in PA of 1.6 degrees.  We find a change of $2.1 \pm 0.9$ degrees.  Evidence ratios comparing best linear fits to best fit constants through all four epochs including uncertainties are 4.5$\times$10$^3$ in PA and 10$^4$ in separation.  The best fit slope of the linear fit in PA is $-0.164 \pm 0.048$ deg/yr and in separation is $0.00058 \pm 0.00032$ arcsec/yr.  We conclude that linear fits are preferred, and that we likely see orbital motion from ROXs 12 b.

Using multiple epochs of astrometry allows us to constrain the orbits of ROXs~42B~b and ROXs~12~b. To fit each orbit we use an updated implementation of the Rejection Sampling Monte Carlo method described in De Rosa et al. (2015), based on the method of Ghez et al. (2008).  This technique generates an initial orbit with semi-major axis ($a$) of unity and position angle of nodes ($\Omega$) of 0, with eccentricity ($e$), inclination angle ($i$), argument of periastron ($\omega$), and epoch of periastron passage ($T_0$) drawn from the appropriate probability distribution: uniform for $e$, $\omega$, T$_0$, and uniform in cos($i$), and we use Kepler's third law to generate the period from a fixed system mass.  We then scale $a$ and rotate $\Omega$ to fit a single observational epoch, with observational errors included by adding Gaussian random noise to the observed separation and position angle for that epoch with $\sigma$ equal to the observational errors.  Stellar mass and distance for each trial are both drawn from Gaussian distributions with medians at the measurements and standard deviations of the measurement uncertainties.  Unlike De Rosa et al. (2015) where all potential orbits where shifted and scaled to the earliest epoch, here we randomly select an epoch for each orbit, which avoids the fit being biased toward the first epoch.

The algorithm has also been modified at the rejection sampling step: previously we proceeded one epoch at a time, rejecting ill-fitting orbits at each epoch.  In this version the chi-square for the newly-scaled orbit is calculated for all the remaining epochs, and then the orbit is accepted if a uniform random variable is less than $e^{-\frac{\chi^2}{2}}$  and rejected otherwise.  Mathematically this is the same operation as we used previously, but it allows for higher computational efficiency in the face of outliers, since the rejection test can be scaled to the minimum value of $\chi^2$ reached for the given astrometry, with orbits now accepted if the random variable is less than $e^{-\frac{\chi^2}{2}}$ / $e^{-\frac{\chi_{min}^2}{2}}$.  This method will be described in more detail in Blunt et al. (2016, in prep).  

This rejection sampling technique produces identical posterior probability distributions to those generated by MCMC, but requires much less computational time for astrometry covering short arcs of an orbit, as demonstrated in De Rosa et al. (2015) for the exoplanet 51 Eri b.  In the very long-period orbits presented here we find that even after $10^{10}$ steps of Metropolis Hastings MCMC the chains have not converged, though the posteriors are broadly similar to those generated by the Rejection Sampling method.

Figures 8 and 9 show the range of Keplerian orbits consistent with the available astrometry for ROXs 12 b and ROXs 42B b respectively, while Figures 10 and 11 show the posterior distributions for the orbital parameters that were fit for ROXs 12 b and ROXs 42B b, respectively.  We note that even with a small fraction of orbital coverage, fitting orbits and obtaining marginal constraints on the corresponding parameters is useful.  For example, several recent studies have fit the small orbital coverage observed for Fomalhaut b, and find that they can constrain the eccentricity of this object to high values \citep{Kalas2013, Beust2015}.  Furthermore, detected orbital motion of the low mass brown dwarfs PZ Tel b and GQ Lup b appears to constrain their eccentricities to high values \citep{Ginski2014}.

\begin{figure*}
\includegraphics[width=1\textwidth]{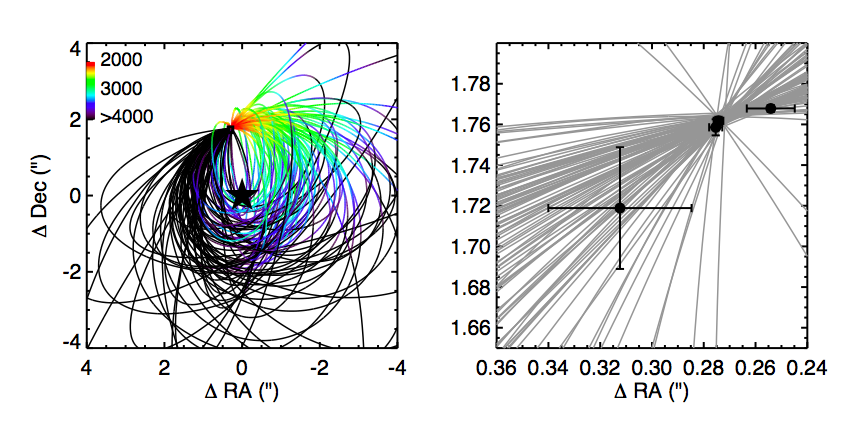}
\caption{100 randomly selected orbital tracks drawn from the posterior distribution for ROXs 12 b.  (left) Colors correspond to elapsed time since 2000.  A clockwise orbit ($i > 90$) is favored, though the astrometric errors allow for a counterclockwise orbit as well.  (right) A zoom-in on the measured astrometry of the system and the same 100 orbital tracks.  Future high-precision astrometric monitoring of the system should improve the constraints on allowable orbits.}
\end{figure*}
\begin{figure*}
\includegraphics[width=1\textwidth]{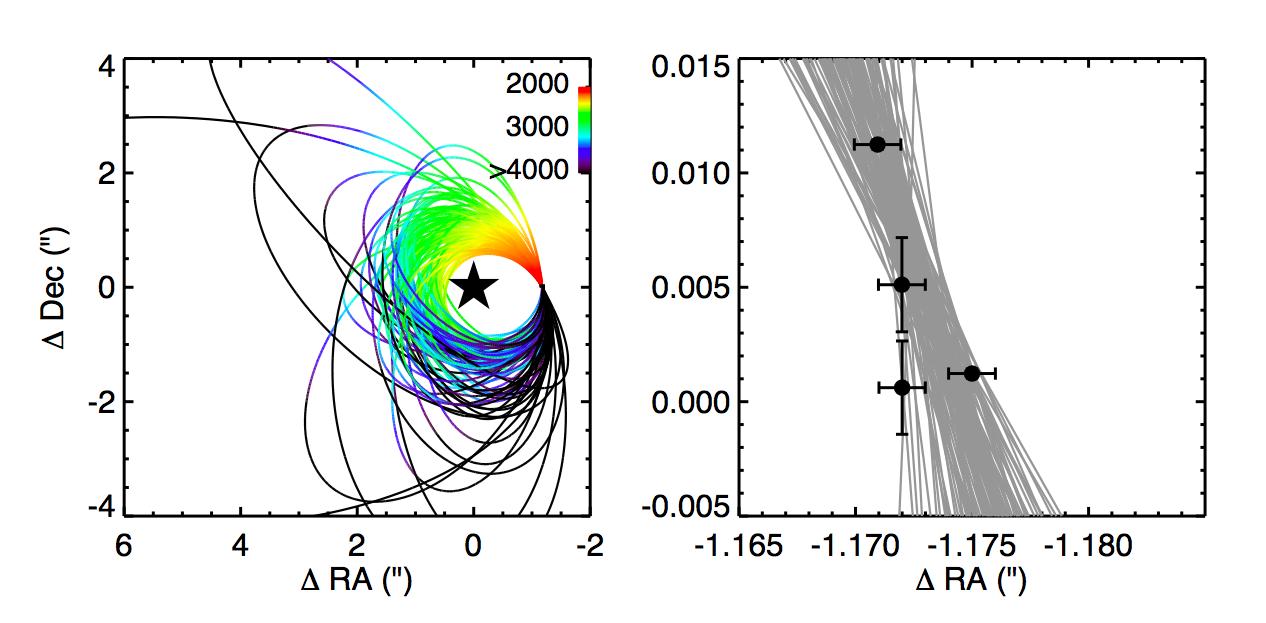}
\caption{Orbital tracks for ROXs 42B b.  See Figure 8 for details.  Generally a face on ($i \lesssim$ 50), circular ($e \lesssim$ 0.5) orbit is preferred.}
\end{figure*}
\begin{figure*}
\includegraphics[width=1\textwidth]{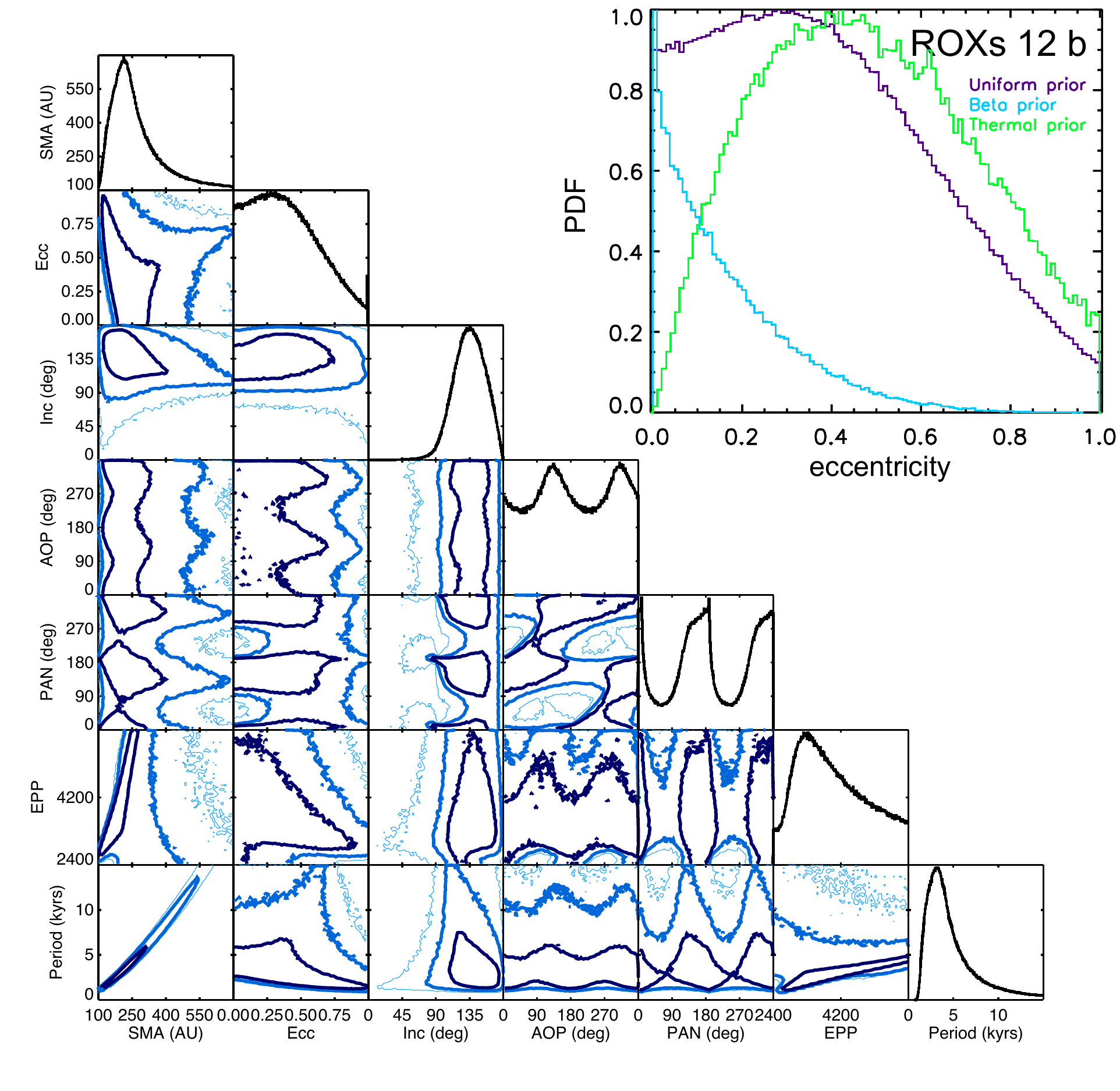}
\caption{The marginalized one-dimensional posterior probability distributions of orbital parameters for ROXs 12 b along the diagonal, and two-dimensional covariances in off-diagonal elements.  Parameters plotted are semi-major axis, eccentricity, inclination, argument of periastron, position angle of nodes, epoch of periastron passage, and period.  In the covariance plots the dark to light blue contours denote locations with 68\%, 95\%, and 99.7\% of the probability enclosed.  The most likely orbits have a semi-major axis of $\sim$200 AU, $\sim$3000 year period, and generally circular ($e \lesssim$ 0.5) and face on ($i \lesssim 70$ or $i \gtrsim$ 110).  In the inset on the upper right, three different eccentricity posteriors are plotted corresponding to three different priors.  The purple, light blue, and dark blue posteriors correspond to a uniform, thermal, and $\beta$ distribution respectively. }
\end{figure*}
\begin{figure*}
\includegraphics[width=1\textwidth]{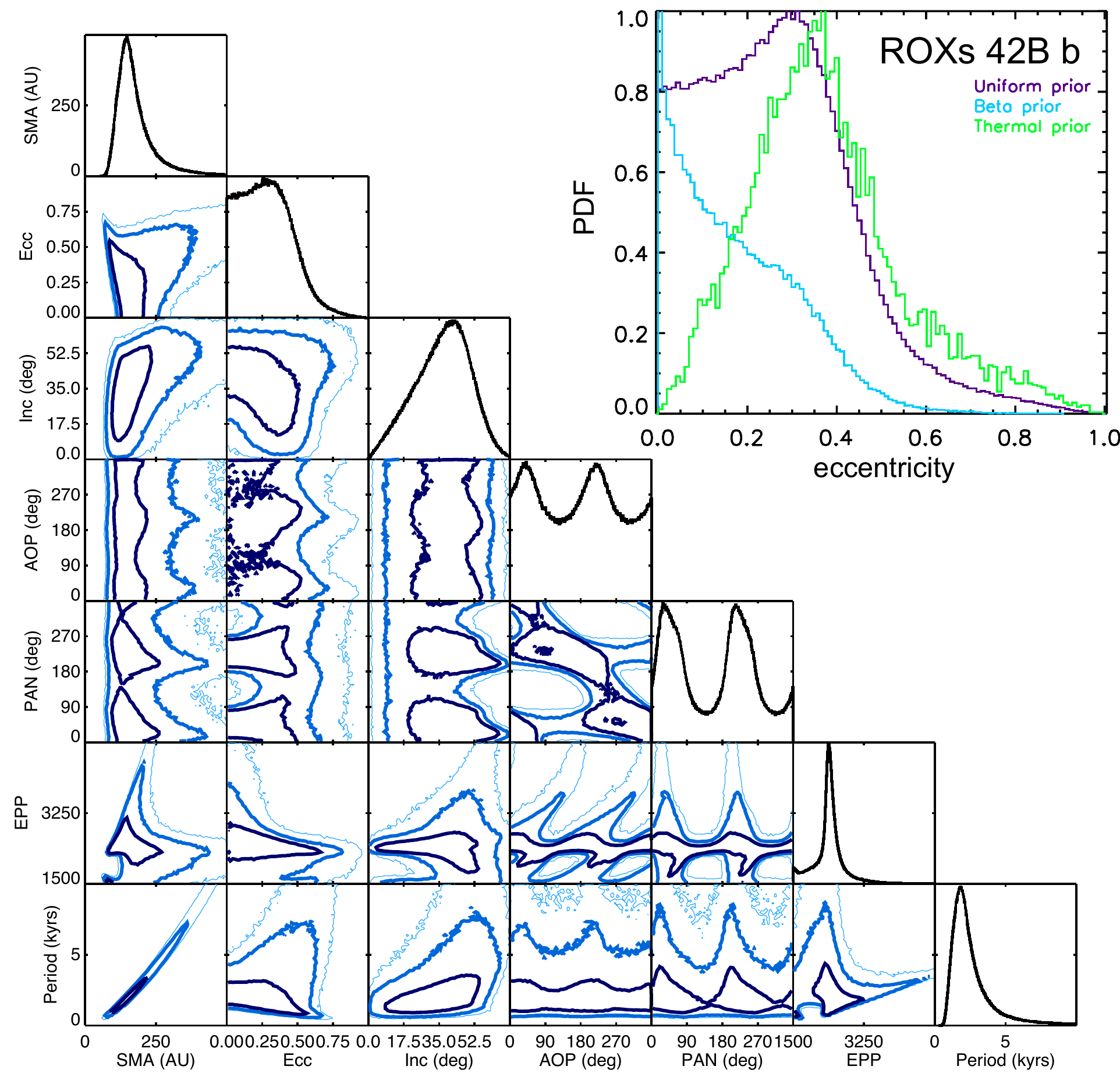}
\caption{Orbital parameter posterior distributions for ROXs 42B. b  The distributions peak for $\sim$150 AU, $\sim$2000 year orbits.  More circular orbits are preferred, with higher inclinations corresponding to longer periods.  As in Figure 10, in the inset on the upper right, three different eccentricity posteriors are plotted corresponding to three different priors.  The purple, light blue, and dark blue posteriors correspond to a uniform, thermal, and $\beta$ distribution respectively.
}
\end{figure*}

While the eccentricities of these PMCs are poorly constrained, we do find that low to moderate eccentricities are favored.  The 95$\%$ upper limits on the eccentricities of ROXs 42B b and ROXs 12 b are 0.58 and 0.83 respectively.  Previous studies have run scattering simulations to test if these wide-separation ($>$ 100 AU) PMCs can form via planet-planet scattering.  These simulations showed that for giant planets that end up outside of 100 AU, their eccentricities are significantly pumped up to $>$ 0.5 \citep{Scharf2009, Nagasawa2011}.  The fact that the eccentricity distributions for ROXs 42B b and ROXs 12 b favor moderate to low eccentricities argues against the scattering hypothesis for these companions.  Note that while a uniform prior on the eccentricity is used in these fits, the eccentricity posterior is significantly different.  We can conclude that the eccentricity posterior is a reflection of the underlying companion eccentricity and not of the prior chosen.  To further test this, we ran these orbit fits with two additional eccentricity priors, the $\beta$ distribution \citep{Kipping2013} and the thermal distribution \citep{Ambartsumian1937}.  The thermal distribution of eccentricities, which is proportional to $2e$ d$e$, is the distribution that binary companions should follow if they are distributed solely as a function of energy.  The eccentricity posteriors using these priors are overplotted with the eccentricity posterior found using a uniform prior in the top right plot in Figures 10 and 11.  For both ROXs 12 b and ROXs 42B b, while the eccentricity posterior using the thermal distribution prior pushes to higher eccentricities, in general lower to moderate eccentricities are favored.

\subsection{Detection Probability}

We calculate the detection probability for additional companions in these eight systems over a range of masses and separations.  Our contrast curves can be converted into sensitivity maps in mass and semi-major axis using evolutionary models, the age and distance of the system and the uncertainties on these values, and an underlying distribution of planet eccentricities.  Following Bowler et al. (2015), we generate a population of artificial companions on random, circular Keplerian orbits with a given mass and semi-major axis.  Each synthetic planet is assigned an apparent magnitude using an interpolated grid of the Cond hot-start evolutionary models (Baraffe et al 2003), the distance and age of the host star, and the companion mass.  We use the Cond evolutionary models because they extend down to planetary masses, although we note that different models can vary significantly in their predictions for the same planet mass.  We do not explicitly account for this model-dependent error in our final analysis.  The fraction of companions falling above a contrast curve compared to those  falling below it yields the fractional sensitivity at that grid point.  We further take into account the fractional field of view coverage for each target, which is uniformly complete out to 4" for our sample and drops to zero beyond that.  Iterating over masses between 0.5--100 M$_{\rm Jup}$ and semi-major axes between 1--1000 AU yields sensitivity maps for each target, which are shown in Figure 12 for this sample.  Depending on the distance and age of the target, our observations are generally sensitive to 1--10~ M$_{\rm Jup}$ companions beyond about 30~AU.

\begin{figure*}
\includegraphics[width=1\textwidth]{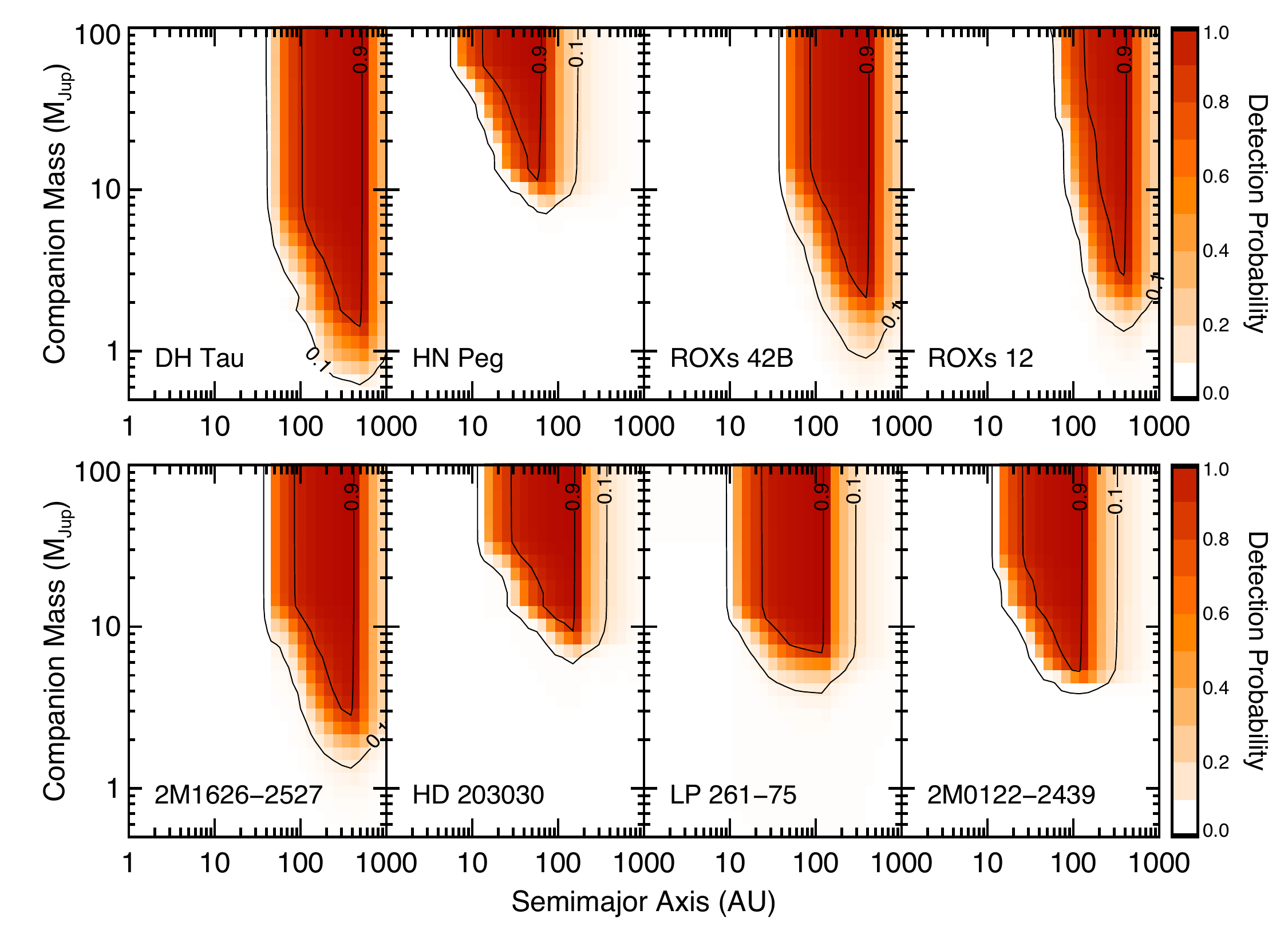}
\caption{Detection probability maps for our sample.  Contours denote the 90$\%$ and 10$\%$ sensitivity regions using the Baraffe et al. (2003) hot-start evolutionary models.  No grid point is exactly 100$\%$ sensitive to companions because even brown dwarfs and low-mass stars at wide orbital distances could be temporarily located at close projected separations from their host star.  The outer drop in sensitivity is caused by limited field of view coverage.  These maps assume circular orbits, but adopting modest eccentricities does not qualitatively change these results.}
\end{figure*}

\section{Discussion}

\subsection{Can Dynamical Scattering Explain This Population?}

Except for DH Tau cc1, which remains ambiguous, none of the new candidate companions detected in our sample are bound.  While we can generally rule out the existence of massive scatterers above our detection limits (outside of 15-50 AU for massive planets), we cannot unambiguously rule out scattering as a formation mechanism based on our results alone, since the massive scatterers might be located closer in.  However, our results combined with complementary lines of evidence suggest that formation close to the host star plus subsequent scattering is probably not the dominant formation mechanism for these wide-separation PMCs.  Note that in order for the scattering scenario to operate, there must be a body that is often at least as massive as these already massive PMCs closer in to the host stars \citep{Veras2004}.  We present a comprehensive list of this evidence below.
\begin{itemize}
\item In this study, we do not find any potential scatterers down to $\sim$ 15 - 50 AU in this sample of seven systems which host wide-separation PMCs.  Furthermore, other studies with comparably deep imaging of wide separation PMCs also did not find any potential scatterers in HD 106906 and 1 RXS 1609-2105 \citep[e.g.][]{Bailey2014, Lafreniere2010, Lagrange2015, Kalas2015}.  Efforts using non-redundant aperture masking techniques have probed higher masses ($>$15 M$_{\rm Jup}$) down to smaller separations ($>$ $\sim 5$ AU), and have likewise found a dearth of inner companions \citep{Cheetham2015, Kraus2011}.  If additional inner gas giant planets or brown dwarfs are present, they must be located within a few tens of AU of their host stars.
\item Moderate to low eccentricities are favored for ROXs 12 b and ROXs 42B b, which both exhibit orbital motion.  This is in contrast to the predictions of scattering simulations, which show that giant planets that get scattered out to $>$100 AU typically have high eccentricities $>$0.5 \citep{Nagasawa2011, Scharf2009}.
\item From RV studies, it is clear that high-mass planets are rare.  This is evident from the significantly negative power law in mass found by \citet{Cumming2008} for a sample of giant planets 0.3 -- 10 M$_{\rm Jup}$ out to 3 AU, where for a power law $m^{\alpha}$, $\alpha = -0.31 \pm 0.2$.   Within this semi-major axis range, this power law implies that the occurrence rate of giant planets in the range 5 - 10 M$_{\rm Jup}$ is $1.3\%$.  Similarly, work by Bryan et al (2016) suggests that for a sample of gas giant planets outside 5 AU, lower-mass planets are more frequent than higher-mass planets.  This implies that planets massive enough to be potential scatterers for the wide-separation directly imaged planets in this study ($>$ 5 M$_{\rm Jup}$) are intrinsically rare in RV surveys.  However, even if massive planets are disfavored generally, this doesn't necessarily mean that there would only be one super-massive planet per system.  For example, perhaps unusual disk properties are required to form a $>$ 5 M$_{\rm Jup}$ planet, but once this kind of disk is formed it is easy to form multiple massive planets at a range of separations.  Given the low estimated occurrence rate of wide-separation PMCs (less than a few percent), this might be consistent with the low occurrence rate of massive planets found in RV surveys.

\item Dynamical interactions between planets preferentially scatter out lower-mass planets \citep{Veras2009}.  However, lower-mass planets ($<$ 5 M$_{\rm Jup}$) have not been discovered at distances greater than 100 AU despite the fact that many surveys were sensitive down to a few Jupiter masses \citep[e.g.][]{Bowler2013,Biller2013}.  This implies that the companion mass function truncates at $\sim$5 M$_{\rm Jup}$, inconsistent with scattering.  
\item  Rough estimates of the occurrence rate of these massive ($>$ 5 M$_{\rm Jup}$) wide-separation PMCs yield at most a frequency of a few percent \citep{Ireland2011, Aller2013}.  In contrast, scattering simulations, which began with 100 systems populated with 10 planets each with masses between 0.1 - 10 M$_{\rm Jup}$ drawn from a uniform distribution in logM and with separations $<$ 30 AU drawn from a uniform distribution in semi-major axis, find that the occurrence rate of scattered planets from 1 -- 10 M$_{\rm Jup}$ outside of 300 AU is $\sim$0.2$\%$ at $\sim$10 Myr \citep{Scharf2009}.  Furthermore, of the 0.2$\%$ occurrence rate of all planets from 1 -- 10 M$_{\rm Jup}$ that get scattered beyond 300 AU, only a few percent of that population are $>$ 4 M$_{\rm Jup}$ \citep{Veras2009}.  This implies that the occurrence rate of massive ($>$ 5 M$_{\rm Jup}$) scattered planets predicted by these simulations is of the order several hundredths of a percent, which is orders of magnitude smaller than any of the occurrence rate measurements from surveys thus far.
\item Many of these widely-separation PMCs are actively accreting from a circumplanetary disk \citep{Zhou2014, Bowler2011}.  However, if these objects were dynamically scattered, we might expect the circumplanetary disk to be partially or completely stripped away \citep{Bowler2011}.  This implies that many if not all of the PMCs that we find did not undergo such a violent evolution and were thus able to keep their disks.

\end{itemize}

Taken together, these lines of evidence indicate that the most likely origin for these wide-separation PMCs is in situ formation.  Evidence for in situ formation, by cloud fragmentation or disk instability, includes the fact that PMCs have been found orbiting low-mass brown dwarfs with decidedly non-planetary mass ratios, implying that the tail of the initial mass function appears to continue down to at least 5 --10 M$_{\rm Jup}$.  In addition, \citet{Brandt2014} found that a single power law distribution is consistent with a sample of 5 -- 70 M$_{\rm Jup}$ objects from the SEEDS survey.  Given that results from many other surveys are well fit by this same power law distribution, this suggests that this population immediately below the deuterium-burning threshold are the end of a smooth mass function, sharing a common origin with more massive brown dwarfs.

\vspace{2cm}

\subsection{Occurrence Rate}
We now consider the multiplicity of directly imaged planetary systems.  Although we did not find any new companions in our sample, we can place an upper limit on the occurrence rate of inner, massive planets in systems with previously known wide-separation PMCs.  Since we have no detections, our occurrence rate is simply:

\begin{equation}
O = \frac{N_{m}}{N_{sys}}
\end{equation}

\noindent where $N_{m}$ is the number of planets that we missed in our survey due to incompleteness, and $N_{sys}$ is the number of systems in our sample. 

The number of planets that we missed due to survey incompleteness can be expressed as:

\begin{equation}
N_{m} =  \Sigma_{i = 0}^{N_{sys}} \bigg[\int_{a_1}^{a_2} d\log a \int_{m_1}^{m_2} d\log m \hspace{0.1cm}f(m,a)(1 - P_i(m,a))\bigg] .
\end{equation} 

\noindent Here, $P_i(m,a)$ is the probability of detecting a planet of mass $m$ at semi-major axis $a$ for system $i$.  We have these values for a grid of masses and semi-major axes from our detection probability calculations, described in section 3.4 and shown in Figure 8.  The quantity $f(m,a)$ is the assumed distribution in mass and semi-major axis for the population of planets whose occurrence rate we wish to calculate.  In our calculation we adopt the underlying distribution in \citet{Clanton2015}, which combines five different exoplanet surveys compiled using three different detection methods to derive a double power law distribution in mass and semi-major axis for giant planets.  This power law takes the form:

\begin{equation}
f(m,a) = \frac{dN}{d\log m_p d\log a} = A \bigg (\frac{m_p}{M_{Sat}} \bigg )^{\alpha}\bigg (\frac{a}{2.5 AU} \bigg )^{\beta}
\end{equation}

\noindent In this equation, $A = 0.21^{+0.20}_{-0.15}$, $\alpha = -0.86^{+0.21}_{-0.19}$, and $\beta = 1.1^{+1.9}_{-1.4}$.  We note that this power law was derived specifically for M dwarf host stars.  Out of our seven systems with previously confirmed PMCs, five of the host stars are M stars.  We then create a 30$\times$30 grid evenly spaced in logarithmic bins with masses from 1 -- 100 M$_{\rm Jup}$ and semi-major axes ranging from 1 -- 1000 AU, and determine the power law distribution values at each grid point.  

Since we want to determine the probability of finding an inner planet given that an outer PMC has been detected, we calculated the occurrence rate of PMCs between 5 -- 15 M$_{\rm Jup}$ and from 40 AU to the location of each PMC.  The inner limit on the separation was chosen because we are reasonably complete for massive planets beyond 40 AU for most of our systems.  In order to take into account the large uncertainties on the power law parameters, we calculated the occurrence rate using a Monte Carlo method with $10^6$ trials, each time drawing a new $A$, $\alpha$, and $\beta$ value from a Gaussian distribution with widths equal to the parameter uncertainties.  This yielded a distribution of missed planets, which we converted to a distribution in occurrence rate.  

We found that the 95$\%$ confidence upper limit on the occurrence rate of planetary companions interior to our sample of previously known wide separation PMCs is 54$\%$.  This result assumes the companion distribution shown in equation 8 as well as hot start evolutionary models.  This first estimate of the occurrence rate upper limit will be better constrained with the discovery and analysis of more PMC systems.  Note that covariances between parameters have not been taken into account in this method, which inflates our upper limit.

\section{Conclusions}

We conducted a deep angular differential imaging (ADI) survey with NIRC2 at Keck in search of close-in substellar companions to a sample of seven systems with confirmed PMCs on extremely wide orbits ($>$100 AU).  We explored the possibility that the wide-separation PMCs formed closer in to their host stars and were subsequently scattered out to their present day locations by a more massive body in the system.  In this survey we obtained deep imaging for each target, for the first time probing significantly lower masses and smaller separations in all systems.  

Within our sample we found eight candidate companions.  Using second epoch data, we measured the astrometry for each candidate and determined whether or not they were co-moving by using an MCMC technique that calculates robust uncertainties from the posterior distributions, without the systematics that occur when the reduced images are used.

Seven candidate companions are unequivocally background objects, while the candidate companion near DH Tau remains ambiguous.  Although our results alone do not conclusively rule out formation closer in to the host star followed by scattering as a formation mechanism for these wide-separation PMCs, the totality of evidence suggests that scattering is not a dominant formation mechanism.  Instead, formation of these objects in situ appears to be more likely.  

If we wish to better understand how these wide separation PMCs formed, there are several possible approaches to consider.  $Gaia$ will allow us to carve out the immediate environment around these young stars, which has been extremely difficult with our current imaging capabilities (due to unfavorable contrasts close to the star), and radial velocity capabilities (due to high jitter values for young stars).  Furthermore, studying the composition of these PMCs by obtaining high resolution spectra might allow us to distinguish amongst formation mechanisms \citep{Konopacky2013, Barman2015}.  While the core accretion model predicts that planets should have enhanced metallicities relative to their host stars, formation via disk instability or turbulent fragmentation should result in compositions matching those of the host star.  Finally, large high-contrast imaging surveys of young star forming regions conducted homogeneously would give us a more precise measurement of the occurrence rates and orbital architectures of this population of planetary-mass objects.

\section*{}

The data presented herein were obtained at the W.M. Keck Observatory, which is operated as a scientific partnership among the California Institute of Technology, the University of California and the National Aeronautics and Space Administration. The Observatory was made possible by the generous financial support of the W.M. Keck Foundation.  We acknowledge the efforts of the Keck Observatory staff.  The authors wish to recognize and acknowledge the very significant cultural role and reverence that the summit of Mauna Kea has always had within the indigenous Hawaiian community.  We are most fortunate to have the opportunity to conduct observations from this mountain.


\begin{thebibliography}{}

\bibitem[Alibert et al(2005)]{Alibert2005}
Alibert, Y., Mordasini, C., Benz, W. et al. 2005, A$\&$A, 434, 1

\bibitem[Aller et al(2013)]{Aller2013}
Aller, K. M., Kraus, A. L., Liu, M. C. et al 2013, ApJ, 773, 1

\bibitem[Ambartsumian(1937)]{Ambartsumian1937}
Ambartsumian, V. A. 1937, Astron. Zh., 14, 207

\bibitem[Bailey et al(2014)]{Bailey2014}
Bailey, V., Meshkat, T., Reiter, M. et al 2014, ApJL, 780, 1

\bibitem[Baraffe et al(2003)]{Baraffe2003}
Baraffe, I., Chabrier, G., Barman, T. S. et al 2003, A$\&$A, 402

\bibitem[Barman et al(2015)]{Barman2015}
Barman, T. S., Konopacky, Q. M., Macintosh, B. et al 2015, ApJ, 804, 1

\bibitem[Bate et al(2002)]{Bate2002}
Bate, M. R., Bonnell, I. A., $\&$ Bromm, V. 2002, MNRAS, 332

\bibitem[Bate(2009)]{Bate2009}
Bate, M. R. 2009, MNRAS, 392

\bibitem[Bate(2012)]{Bate2012}
Bate, M. R. 2012, MNRAS, 419

\bibitem[Beust et al(2015)]{Beust2015}
Beust, H., Bonnefoy, M., Maire, A. L. et al 2015, A$\&$A, accepted

\bibitem[Biller et al(2013)]{Biller2013}
Biller, B. A., Liu, M. C., Wahhaj, Z. et al 2013, ApJ, 777, 2

\bibitem[Boss(2006)]{Boss2006}
Boss, A. P. 2006, ApJ, 637

\bibitem[Bowler et al(2011)]{Bowler2011}
Bowler, B. P., Liu, M. C., Kraus, A. L. et al 2011, ApJ, 743, 2

\bibitem[Bowler et al(2013)]{Bowler2013}
Bowler, B. P., Liu, M. C., Shkolnik, E. L. et al 2013, ApJ, 774, 1

\bibitem[Bowler et al(2014)]{Bowler2014}
Bowler, B. P., Liu, M. C., Kraus, A. L. et al 2014, ApJ, 784, 1

\bibitem[Bowler et al(2015)]{Bowler2015}
Bowler, B. P., Shkolnik, E. L., Liu, M. C. et al 2015, ApJ, 806, 1

\bibitem[Bowler et al(2015)]{Bowler20152}
Bowler, B. P., Liu, M. C., Shkolnik, E. L. et al 2015, ApJS, 216, 1


\bibitem[Brandt et al(2014)]{Brandt2014}
Brandt, T. D., McElwain, M. W., Turner, E. L. et al 2014, ApJ, 794, 2

\bibitem[Bryan et al(2016)]{Bryan2016}
Bryan, M. L., Knutson, H. A., Howard, A. W. et al 2016, ApJ, 821, 89

\bibitem[Carlsberg Meridian Catalogue 15(2011)]{CMC2011}
Carlsberg Meridian Catalogue 15, 2011

\bibitem[Chauvin et al(2004)]{Chauvin2004}
Chauvin, G., Lagrange, A. M., Lacombe, F. et al 2004, A$\&$A, 425

\bibitem[Chauvin et al(2005)]{Chauvin2005}
Chauvin, G., Lagrange, A. M., Zuckerman, B. et al 2005, A$\&$A, 438, 3

\bibitem[Cheetham et al(2015)]{Cheetham2015}
Cheetham, A. C., Kraus, A. L., Ireland, M. J. et al 2015, ApJ, 813, 2

\bibitem[Clanton $\&$ Gaudi(2015)]{Clanton2015}
Clanton, C. $\&$ Gaudi, S. 2015, arXiv:150804434C

\bibitem[Cumming et al(2008)]{Cumming2008}
Cumming, A., Butler, R. P., Marcy, G. W. et al. 2008, PASP, 120, 531

\bibitem[Currie et al(2015)]{Currie2015}
Currie, T., Cloutier, R., Brittain, S. et al 2015, ApJL, 814, 2

\bibitem[Currie et al(2014)]{Currie2014}
Currie, T., Daemgen, S., Debes, J. et al 2014, ApJ, 780, 2

\bibitem[Cutri et al(2003)]{Cutri2003}
Cutri, R. M., Skrutskie, M. F., van Dyk, S. et al 2013, 2MASS All Sky Catalog

\bibitem[Cutri et al(2013)]{Cutri2013}
Cutri, R. M., Wright, E. L., Conrow, T. et al 2013, AllWISE Data Release

\bibitem[De Rosa et al(2015)]{DeRosa2015}
De Rosa, R. J., Nielsen, E. L., Blunt, S. C. et al 2015, ApJ, 814, 3

\bibitem[Dodson-Robinson et al(2009)]{Dodson2009}
Dodson-Robinson, S. E., Veras, D., Ford, E. B. et al 2009, ApJ, 707

\bibitem[Dupuy $\&$ Liu(2011)]{Dupuy2011}
Dupuy, T. J. $\&$ Liu, M. C. 2011, ApJ, 733, 2

\bibitem[Ghez et al(2008)]{Ghez2008}
Ghez, A. M., Salim, S., Weinberg, N. N. et al 2008, ApJ, 689, 1044

\bibitem[Ginski et al(2014)]{Ginski2014}
Ginski, C., Schmidt, T. O. B., Mugrauer, M. et al 2014, MNRAS, 444, 3

\bibitem[Ireland et al(2011)]{Ireland2011}
Ireland, M. J., Kraus, A., Martinache, F. et al 2011, ApJ, 726, 2

\bibitem[Itoh et al(2005)]{Itoh2005}
Itoh, Y., Hayashi, M., Tamura, M. et al 2005, ApJ, 620, 2

\bibitem[Kalas et al(2015)]{Kalas2015}
Kalas, P. G., Rajan, A., Wang, J. J. et al 2015, ApJ, 814, 1

\bibitem[Kalas et al(2013)]{Kalas2013}
Kalas, P., Graham, J. R., Fitzgerald, M. P. et al 2013, ApJ, 775, 1

\bibitem[Kipping(2013)]{Kipping2013}
Kipping, D. M. 2013, MNRAS, 434, L51

\bibitem[Konopacky et al(2013)]{Konopacky2013}
Konopacky, Q. M. et al 2013, Science, 229, 1398

\bibitem[Kraus et al(2014)]{Kraus2014}
Kraus, A. L., Ireland, M. J., Cieza, L. A. et al 2014, ApJ, 781, 1

\bibitem[Kraus et al(2011)]{Kraus2011}
Kraus, A. L., Ireland, M. J., Martinache, F. et al 2011, ApJ, 731, 1

\bibitem[Kraus $\&$ Ireland(2012)]{Kraus2012}
Kraus, A. $\&$ Ireland, M. J. 2012, ApJ, 745, 1 

\bibitem[Lafreniere et al(2010)]{Lafreniere2010}
Lafreniere, D., Jayawardhana, R., $\&$ van Kerkwijk, M. H. 2010, ApJ, 719, 1

\bibitem[Lagrange et al(2015)]{Lagrange2015}
Lagrange, A. M., Langlois, M., Gratton, R. et al 2015, arXiv:  1510.02511

\bibitem[Lambrechts $\&$ Johansen(2012)]{Lambrechts2012}
Lambrechts, M. $\&$ Johansen, A. 2012, A$\&$A, 544

\bibitem[Lodato et al(2005)]{Lodato2005}
Lodato, G., Delgado-Donate, E., $\&$ Clarke, C. J. 2005, MNRAS, 364, 1

\bibitem[Low $\&$ Lynden-Bell(1976)]{Low1976}
Low, C. $\&$ Lynden-Bell, D. 1976, MNRAS, 176

\bibitem[Luhman et al(2007)]{Luhman2007}
Luhman, K. L., Patten, B. M., Marengo, M. et al 2007, ApJ, 654, 1

\bibitem[Luhman et al(2006)]{Luhman2006}
Luhman, K. L., Wilson, J. C., Brandner, W. et al 2006, ApJ, 649, 2


\bibitem[Marois et al(2008)]{Marois2008}
Marois, C., Macintosh, B., Barman, T. et al 2008, Science, 322, 5906

\bibitem[Marois et al(2010)]{Marois2010}
Marois, C., Machintosh, B., $\&$ Veran, J. P. 2010, Proceedings of the SPIE, 7736

\bibitem[Metchev $\&$ Hillenbrand(2006)]{Metchev2006}
Metchev, S. A. $\&$ Hillenbrand, L. A. 2006, ApJ, 651, 2

\bibitem[Nagasawa $\&$ Ida(2011)]{Nagasawa2011}
Nagasawa, M. $\&$ Ida, S. 2011, ApJ, 742, 2

\bibitem[Pollack et al(1996)]{Pollack1996}
Pollack, J. B., Hubickyj, O., Bodenheimer, P. et al 1996, Icarus, 124, 1

\bibitem[Quanz et al(2015)]{Quanz2015}
Quanz, S. P., Amara, A., Meyer, M. R. et al 2015, ApJ, 807, 1

\bibitem[Ratzka et al(2005)]{Ratzka2005}
Ratzka, T., Kohler, R., $\&$ Leinert, C. 2005, A$\&$A, 437, 2

\bibitem[Reid $\&$ Walkowicz(2006)]{Reid2006}
Reid, I. N. $\&$ Walkowicz, L. M. 2006, PASP, 118, 843

\bibitem[Sallum et al(2015)]{Sallum2015}
Sallum, S., Follette, K. B., Eisner, J. A. et al 2015, Nature, 527, 7578

\bibitem[Scharf $\&$ Menou(2009)]{Scharf2009}
Scharf, C. $\&$ Menou, K. 2009, ApJ, 693

\bibitem[Service et al.(2016)]{Service}
Service, M., Lu, J. R., Campbell, R. et al 2016, submitted

\bibitem[Skiff(2013)]{Skiff2013}
Skiff, B. A. 2013, General Catalogue of Stellar Spectral Classifications

\bibitem[Soummer et al(2012)]{Soummer2012}
Soummer, R., Pueyo, L., $\&$ Larkin, J. 2012, ApJL, 755, 2

\bibitem[Strom et al(1989)]{Strom1989}
Strom, K. M., Strom, W. E., Edwards, S. et al 1989, AJ, 97, 1451

\bibitem[Veras et al(2009)]{Veras2009}
Veras, D., Crepp, J. R., Ford, E. B. 2009, ApJ, 696

\bibitem[Veras $\&$ Armitage(2004)]{Veras2004}
Veras, D. $\&$ Armitage, P. J. 2004, MNRAS, 347, 2


\bibitem[Vorobyov(2013)]{Vorobyov2013}
Vorobyov, E. I. 2013, A$\&$A, 552

\bibitem[White $\&$ Ghez(2001)]{White2001}
White, R. J. $\&$ Ghez, A. M. 2001, ApJ, 556, 265

\bibitem[Yelda et al(2010)]{Yelda2010}
Yelda, S., Lu, J. R., Ghez, A. M. et al 2010, ApJ, 725, 1

\bibitem[Zacharias et al(2005)]{Zacharias2005}
Zacharias, N., Monet, D. G., Levine, S. E. et al 2005, NOMAD Catalog

\bibitem[Zacharias et al(2012)]{Zacharias2012}
Zacharias, N., FInch, C. T., Girard, T. M. et al 2012, UCAC4

\bibitem[Zhou et al(2014)]{Zhou2014}
Zhou, Y., Herczeg, G. J., Kraus, A. L.  et al 2014, ApJL, 783, L17

 
\end{thebibliography}
\end{document}